\documentclass[preprintnumbers,amsmath,amssymb,floatfix,11pt,prd,onecolumn,superscriptaddress,nofootinbib]{revtex4}
\usepackage[utf8]{inputenc}
\usepackage{diagbox}
\usepackage{latexsym}
\usepackage{epsfig}
\usepackage{epstopdf}
\usepackage{float}
\usepackage{graphicx}
\usepackage{amssymb}
\usepackage{amsmath}
\usepackage{dcolumn}
\usepackage{subfigure}
\usepackage[caption=false]{subfig}
\usepackage{bm}
\usepackage{color}
\usepackage{comment}
\usepackage[shortlabels]{enumitem}
\usepackage{subcaption}
\usepackage{multirow}
\usepackage{xcolor}
\begin{document}
\title{\bf Impact of a Cold Dark Matter Halo on Magnetic Reconnection and Energy Extraction from Kerr-like Black Holes}

\author{Muhammad Nawaz}
\email{mnawazzz158@gmail.com}
\affiliation{Department of Mathematics, University of Okara, Okara-56300, Pakistan}

\author{Abdul Malik Sultan}
\email{ams@uo.edu.pk, maliksultan23@gmail.com}
\affiliation{Department of Mathematics, University of Okara, Okara-56300, Pakistan}

\author{Muhammad Israr Aslam}
\email{mrisraraslam@gmail.com, israr.aslam@umt.edu.pk}
\affiliation{Department of Mathematics, School of Science, University of Management and Technology, Lahore-$54770$, Pakistan}

\author{Rabia Saleem}
\email{rabiasaleem@cuilahore.edu.pk}
\affiliation{Department of Mathematics, COMSATS University Islamabad, Lahore Campus, Lahore-$54000$, Pakistan}

\author{Ke Wang}
\email{kkwwang2025@163.com}
\affiliation{School of Material Science and Engineering, Chongqing Jiaotong University, Chongqing 400074, China}

\begin{abstract}
Recently, Comisso and Asenjo introduced a new energy extraction mechanism based on magnetic reconnection. In this paper, we investigate the power and efficiency of magnetic reconnection energy extraction in a rotating black hole surrounded by a cold dark matter halo. We first examine the properties of the underlying spacetime and its physical quantities, including the event horizon, photon sphere, and ergosphere. We then analyze the allowed energy extraction region and the corresponding energy extraction efficiency for circular orbits. Our results show that energy extraction is feasible for black holes with a spin parameter as low as $0.86$. We further demonstrate that the extracted power exceeds that of the Blandford–Znajek mechanism. In addition, we find that the dark matter halo parameters, $\rho_c$ and $R_s$, reduce the minimum black hole spin required for energy extraction. We also investigate the energy extraction region, power, and efficiency in the plunging region. Our analysis shows that the minimum spin required for energy extraction decreases to $a=0.25$ in the plunging region. As in the circular orbit case, the parameters $\rho_c$ and $R_s$ further reduce the minimum spin required for energy extraction. Finally, we compare the extracted power in the plunging and circular orbit regions and find that the power in the plunging region is greater than that in the circular orbit case.
\end{abstract}
\maketitle
\section{INTRODUCTION }
Black holes are among the most fascinating and significant astrophysical objects in the universe. Since their prediction by Einstein's theory of general relativity (GR), they have played a fundamental role in modern physics and astrophysics. Compelling observational evidence for the existence of black holes emerged with the detection of gravitational waves by the Laser Interferometer Gravitational-Wave Observatory (LIGO) \cite{abbott2016observation,abbott2016gw151226,abbott2017gw170104,abbott2017gw170814} and the first imaging of a black hole shadow by the Event Horizon Telescope (EHT) \cite{akiyama2019first,event2022first}. These groundbreaking discoveries have established black holes as one of the most active and influential research topics in modern astrophysics. Despite these remarkable achievements, many aspects of black hole physics remain unexplored, and further observational and theoretical studies are expected to reveal new insights into their nature and fundamental properties.

As one of the most compact objects with an extremely strong gravitational field, a black hole is closely associated with some of the most energetic astrophysical phenomena in the universe, including active galactic nuclei (AGNs) \cite{mckinney2004measurement,hawley2006magnetically,komissarov2007meissner,tchekhovskoy2011efficient} and gamma-ray bursts (GRBs) \cite{lee2000blandford,tchekhovskoy2008simulations,komissarov2009activation}. It is generally believed that such an enormous amount of energy associated with black holes, including gravitational energy and the electromagnetic energy generated in the vicinity of the black hole, can be produced by the black hole itself. Understanding the possible mechanisms of energy extraction is essential for revealing the nature of the observed astrophysical phenomena powered by black holes.
The studies of Christodoulou demonstrated that, for a Kerr  black hole with mass $M$ and spin parameter $a$, a portion of its mass
\begin{equation}
 M_{irr} =M\sqrt{\frac{1}{2}(1+\sqrt{1-\frac{a^2}{M^2}})},
 \label{mass of BH}
\end{equation}
which is associated with the horizon entropy $S=4\pi M^2_{irr}$ , is reducible. According to the second law of black hole thermodynamics, the maximum amount of energy that can be extracted from an extremal Kerr black hole reaches approximately $E_{rot}\approx 0.29M$, corresponding to the case $a/M = 1$. This establishes the maximum amount of extractable rotational energy, leaving the fundamental question of how to efficiently extract such energy from a rotating black hole. A key feature of a rotating black hole is the existence of the ``ergosphere'', which is defined as the region where the asymptotic time-translation killing field $\xi^a=(\partial/\partial t)^a$ becomes space-like. This condition is characterized by $\xi^a\xi_a=g_{tt}>0$ , indicating that the corresponding tangent vector is space-like. A remarkable consequence is that, within the ergosphere, the energy of a test particle, as measured by a distant observer, can become negative.
\par Exploiting this unique property, Roger Penrose proposed the first theoretical mechanism for extracting energy from a rotating black hole. This mechanism subsequently became well known in the scientific community as the Penrose process. The Penrose process is based on a thought experiment involving particle fission. Specifically, an incident particle $0$ enters the ergosphere and further splits into two particles, namely particle $1$ and particle $2$ ( i.e., $0\rightarrow 1+2$ ). Particle $1$ falls into the black hole, whereas particle $2$ escapes from the black hole and propagates to infinity. Under normal circumstances, particle $2$ would be expected to possess less energy than the initial particle (particle $0$), from which it originated. Therefore, for a distant observer, particle $1$ can acquire negative energy. As a consequence, particle $2$ may escape to infinity with an energy greater than that of the initial particle. When the black hole absorbs-negative energy particle, its mass and spin decrease. Consequently, a net amount of energy can be extracted from the black hole as observed by a distant observer, causing the black hole to spin down. This process can, in principle, be repeated until all the rotational energy of the black hole has been extracted, eventually leaving the black hole non-rotating state.
\par Since particles $1$ and $2$ must separate with extremely high relative velocities, and the occurrence of such particle fission events is exceedingly rare \cite{banados2009kerr,wei2010charged}, the original Penrose process is expected to be highly inefficient for extracting the rotational energy of a black hole. Despite its limitations, the Penrose process established the foundation for the development of more efficient energy extraction mechanisms. Subsequently, several improved mechanisms were proposed, including the particle accelerator mechanism \cite{banados2009kerr,wei2010charged}, superradiant scattering \cite{teukolsky1974perturbations}, the collisional Penrose process \cite{piran1975high}, the Blandford–Znajek process \cite{blandford1977electromagnetic}, and the magnetohydrodynamic process \cite{takahashi1990magnetohydrodynamic}. Similarly, the repetitive Penrose process in Rastall rotating black holes immersed in quintessence dark energy has been investigated in \cite{sabir1}. Among these, the Blandford–Znajek process, which exploits the surrounding magnetic field, has been regarded as the most promising mechanism for powering the relativistic jets of AGNs and  GRBs.

Large-scaleale magnetic fields are expected to exist in the vicinity of black holes and play a crucial role in energy extraction processes in various astrophysical environments. In particular, the energy extraction mechanism through magnetic reconnection, proposed by Comisso and Asenjo \cite{comisso2021magnetic}, has attracted considerable attention in recent years. Fast magnetic reconnection creates a current sheet that converts magnetic energy into the kinetic energy of plasma particles, allowing them to escape through the reconnection outflows \cite{daughton2009transition,bhattacharjee2009fast}. During this process, one portion of the plasma is accelerated to high velocities, while the other portion is correspondingly decelerated. We assume that the accelerated plasma particles co-rotate with the black hole and escape to infinity, whereas the decelerated particles move toward the black hole and are eventually absorbed by it. Similar to the Penrose process, magnetic reconnection takes place within the ergosphere, where the decelerated plasma particles can possess negative energy as measured by a distant observer. Consequently, according to the law of energy conservation, the accelerated plasma particles escape carrying more energy. As the plasma particles leave the magnetic reconnection region, the magnetic tension causes the magnetic field lines to contract again. Owing to the frame-dragging effect of the rotating black hole, the magnetic field lines are continuously twisted, allowing the fast magnetic reconnection process to occur repeatedly. Consequently, this energy extraction mechanism can operate continuously around rapidly rotating black holes. For this process to be effective, the reconnection X-point must lie within the ergosphere, ensuring that the decelerated plasma particles acquire negative energy as measured by a distant observer.
\par Recently, Comisso and Asenjo achieved a significant breakthrough by investigating the efficiency and power of the magnetic reconnection mechanism for energy extraction in the background of a Kerr black hole. Their results demonstrated that, within certain regions of the parameter space, the magnetic reconnection mechanism can be more efficient than the Blandford–Znajek process. Following the work of Comisso and Asenjo, the magnetic reconnection mechanism for black hole energy extraction has been extensively explored in a variety of black hole spacetimes, including spinning braneworld black holes \cite{wei2022effects}, non-Kerr black holes \cite{israr1}, black holes with and without Lorentz symmetry \cite{khodadi2022magnetic}, Lorentz breaking Kerr-Sen and Kiselev black holes \cite{carleo2022energy}, and several other black hole models \cite{wang2022extracting,li2023energy,l2023energy,zhang2024energy,zhang2024,khodad2023harvesting,shaymatov2024kerr,rodriguez2025energy,long2025magnetic,zeng2025energy,wang2025energy,zeng2025e2,eshtursunov2026energy,eshtursunov2026magnetic,yao2026energy,enhanced,yuchih2025energy,cheng2025extractin,shen2024energy,nawaz1}. These studies consistently demonstrate that parameters such as the tidal charge, the Lorentz breaking parameter, and other black hole parameters significantly influence the efficiency of energy extraction through the magnetic reconnection mechanism. Also, such mechanism may dominate in extracting black hole rotational energy.
\par In cosmology, the study of the large-scale structure, evolution, and dynamics of the universe has emerged as one of the most active and rapidly developing fields of modern research. Dark matter (DM) halos, which are gravitationally bound concentrations of DM, play a fundamental role in the complex nonlinear process of cosmic structure formation. These halos serve as the gravitational framework within which luminous matter accumulates, thereby governing the formation and evolution of galaxies, galaxy clusters, and superclusters. Consequently, understanding the physical properties and dynamical behaviour of DM halos is essential for explaining the formation of large-scale structures and for establishing consistency with cosmological structure formation models \cite{alimi2024shape}. Furthermore, according to Newtonian mechanics, the rotational velocity of stars is given by $v^{2}=GM/r$, implying that the orbital velocity of stars should decrease with increasing distance from the galactic centre. However, astronomical observations have shown that the rotation curves of galaxies, including that of the Milky Way, remain nearly flat even at distances far beyond the visible edge of the galactic disk \cite{bottema1997investigation}. This discrepancy led to the prediction of the existence of DM, which has since been supported by a wide range of indirect observational evidence \cite{gammaldi2018theoretical}. As a result, extensive research has been devoted to DM, leading to the development of several theoretical models, including the Cold Dark Matter (CDM) model \cite{navarro1997universal}, the Warm Dark Matter (WDM) model \cite{bode2001halo}, and others. Among these cosmological models, the CDM model is of particular importance, contributing approximately $26\%$ of the total energy density of the Universe \cite{planck2020cmb}. According to the Navarro–Frenk–White (NFW) profile, the density distribution of a DM halo in the CDM model is expressed as \cite{navarro1996structure,navarro1995simulations}
\begin{equation} 
\rho_{\rm NFW}=\frac{\rho_c}{\left(r/R_s\right)\left(1+r/R_s\right)^2},
\label{densitydistribution}
\end{equation}
where $\rho_c$ represents the critical density and $R_s$ is the scale radius of the DM halo.
\par The relationship between black holes and  DM has become an important topic of research in the scientific community. Several studies have suggested that the presence of DM can give rise to a density spike in the vicinity of a black hole \cite{sadeghian2013dark} while DM itself may also facilitate the formation of black holes \cite{zelnikov2005influence}. Observational evidence indicates that supermassive black holes located at the centres of galaxies are surrounded by DM particles \cite{lynden1969galactic,gebhardt2011black}. 
In this context, the thermodynamic properties of a rotating black hole immersed in a CDM halo were investigated in \cite{majhi2024thermodynamics}, where it was found that stable black hole solutions exist only for relatively small values of the characteristic density parameter $\rho_c$. In Ref. \cite{meng2024destroying}, the authors examined the weak cosmic censorship conjecture in a more realistic astrophysical scenario by considering rotating black holes surrounded by DM. Furthermore, the motion of particles orbiting a black hole embedded in a DM halo was studied in \cite{tan2025motion}. The authors concluded that the presence of a DM halo can significantly modify the orbital trajectories, particle energies, and the parameters of the innermost stable circular orbit (ISCO). In general, astrophysical black holes are not isolated compact objects. Their accretion disks are embedded within DM halos, which may extend throughout the host galaxy, including the Milky Way. Photons emitted from the accreting matter are also gravitationally lensed by the surrounding DM halo, thereby enhancing the observed optical appearance of the black hole \cite{macedo2024optical}. Recently, Zeng et al., \cite{sadia} discussed the shadow images of a Kerr-like black hole in CDM halo illuminated with a celestial light source and a thin accretion disk.

\par In Ref. \cite{yang2024pseudoisothermal}, the authors investigated the influence of a pseudo-isothermal DM halo on the event horizon, time-like and null geodesics, and the shadow of a black hole. Their analysis showed that the DM density profile has only a negligible effect on the shadow of the black hole in $M87$. Considering the effects of the Dehnen DM halo profile, a rotating black hole solution was constructed using the Newman–Janis algorithm in Ref. \cite{pantig2022dehnen}. The obtained results indicated that, although dwarf galaxies are rich in DM, the influence of the Dehnen profile on the black hole depends strongly on its mass. The shadow and deflection angle of a black hole in the presence of the Dekel–Zhao DM profile were investigated in Ref. \cite{ovgun2025dekelzhao}. The results demonstrated that the shadow radius increases with the black hole mass, whereas higher central DM densities lead to a smaller shadow, highlighting the environmental effects of dense DM halos. Pantig and Övgün \cite{pantig2023quantum} studied the effects of a fuzzy (or wave) DM halo surrounding a supermassive black hole. They examined the behaviour of the shadow radius under observational constraints from the perspective of a static observer and further discussed the influence of the soliton profile on a thin accretion disk. In Ref. \cite{liu2023gravitational}, the authors investigated the gravitational ringing of black holes in low surface brightness (LSB) galaxies within the CDM and scalar field DM models and compared their results with the Kerr black hole case. By analysing the frequencies of quasinormal modes (QNMs) and quasi-bound states (QBSs) for different parameter values, they confirmed the existence of superradiant instability in both DM models. The effects of three DM density profiles, namely the CDM, scalar field DM, and Burkert profiles, on the weak deflection angle derived from the universal rotation curve were examined in Ref. \cite{pantig2022weak}. Their results showed that variations in the weak deflection angle depend strongly on both the DM mass and the adopted density profile. Furthermore, four spherically symmetric but non-asymptotically flat black hole solutions surrounded by a spherical DM distribution were obtained under the influence of a minimal length through the generalized uncertainty principle in Ref. \cite{ovgun2024constraints}. Within this framework, the effects of the quantum correction parameter ($\gamma$) on the event horizon, photon sphere, and shadow radius were analyzed, revealing significant deviations from the Schwarzschild black hole case. Motivated by these developments, the present work focuses on the astrophysical implications of rotating black holes immersed in a CDM halo, with particular emphasis on their observational properties and energy extraction through the magnetic reconnection mechanism.
\par In this paper, we investigate the mechanism of energy extraction via magnetic reconnection in a rotating black hole surrounded by a CDM halo. Our primary objective is to examine how the presence of the CDM halo influences the efficiency of the magnetic reconnection process and to analyze its impact on the energy extraction from the black hole. We also examine the energy can be extracted through the magnetic reconnection mechanism even from moderate spin black holes, thereby extending the applicability of the mechanism. This improvement is primarily attributed to the CDM parameters $\rho_c$ and $R_s$, which reduce the minimum black hole spin required for energy extraction and make the magnetic reconnection process possible for more moderately rotating black holes.

\par This paper is organized as, In section {\bf II}, we briefly define the background of the black hole model surrounded by CDM halo and discuss its geometrical and physical properties. In section {\bf III. A}, we describe the magnetic reconnection process in circular orbits. In section {\bf III.B}, we analyzes the allowed parameter space for energy extraction. While section {\bf III.C} is devoted to discussing the corresponding power and efficiency of energy extraction. In section {\bf IV.A}, we present the magnetic reconnection mechanism in the plunging region, while in section {\bf IV.B}, we analyze the associated power  of energy extraction. Finally, in the last section we put  our conclusions and discussions.

\section{A Brief Review Of The Rotating Black Hole surrounded by cold dark matter halos} 

We begin by introducing the underlying background geometry. The presence of DM halos in galaxies can be characterized by several physical properties, including the scale radius and the characteristic density arising from particle interactions. When a black hole is located at the centre of a DM halo, it can interact with the surrounding DM. The black hole-DM system can be accurately described by adopting the steady state approximation for the black hole spacetime. As described in Eq. (\ref{densitydistribution}), the spacetime metric of a rotating black hole immersed in a CDM halo can be expressed as follows \cite{majhi2024thermodynamics,sadia}:
\begin{equation}
\begin{aligned} ds^{2}={}&-\left(1-\frac{r^{2}+a^{2}-\Delta}{\hat{\Sigma}^{2}}\right)dt^{2} +\left(r^{2}+a^{2}-\Delta\right)\,dt\,d\phi \\ &+\frac{\hat{\Sigma}^{2}}{\Delta}\,dr^{2} +\hat{\Sigma}^{2}\,d\theta^{2} +\frac{\left(r^{2}+a^{2}\right)^{2}-a^{2}\Delta\sin^{2}\theta} {\hat{\Sigma}^{2}} \sin^{2}\theta\,d\phi^{2}, \end{aligned}
\label{metric}
\end{equation}
along with
\begin{equation}
\Delta=a^{2}-2Mr+r^{2}\left(1+\frac{r}{R_{s}}\right)^{-\frac{8\pi\rho_{c}R_{s}^{3}}{r}},\qquad \hat{\Sigma}^{2}=r^{2}+a^{2}\cos^{2}\theta,
\label{components}
\end{equation}
where $M$ and $a$ corresponds to the mass and spin parameter of the black hole, respectively. In the limit of $\rho_c \rightarrow 0$, the metric \ref{metric} reduces to Kerr black hole metric. Further, by substituting $\Delta=0$, one can find the horizons of black hole surrounded by CDM halos. This equation give two positive roots, larger called event horizon $r_+$ and smaller called Cauchy horizon $r_-$. And $g_{tt}=0$ gives the boundary of ergosphere, which give one root that lie outside the event horizon.
\par Now we discuss the motion of photons around the rotating black hole in the CDM halo. Since photons follow null geodesics in the given black hole spacetime, the geodesic motion for the spacetime given by  (\ref{metric}) is governed by the Hamilton-Jacobi equation \cite{chandrasekhar1998mathematical},
\begin{equation}
\frac{\partial I}{\partial \lambda}=-\frac{1}{2}g^{\mu\nu}\frac{\partial I}{\partial x^\mu}\frac{\partial I}{\partial x^\nu}.
\label{jacobiequation}
\end{equation}
where $\lambda$ is the affine parameter and $I$ is the Jacobi action of the photon. The Jacobi action can be separated as
\begin{equation}
I=\frac{1}{2}\varepsilon^{2}\lambda-\hat{E}t+\hat{L}\phi+B_r(r)+B_\theta(\theta),
\label{jacobiaction}
\end{equation}
where $\varepsilon=0$ for photons. The constants $\hat{E}=-p_t$ and $\hat{L}=p_\phi$ denote the conserved energy and conserved angular momentum of the photon, respectively. The functions $B_r(r)$ and $B_\theta(\theta)$ depend only on $r$ and $\theta$. Substituting Eq.~\ref{jacobiaction} into Eq.~\ref{jacobiequation}, one obtains the following equations of motion:
\begin{align}
\hat{\Sigma^2}\frac{dt}{d\lambda}&=a\left(\hat{L}-a\hat{E}\sin^{2}\theta\right)+\frac{r^{2}+a^{2}}{\Delta}\left[\hat{E}(r^{2}+a^{2})-a\hat{L}\right],  \\ \hat{\Sigma^2}\frac{dr}{d\lambda}&=\pm\sqrt{R(r)}, \\ \hat{\Sigma^2}\frac{d\theta}{d\lambda}&=\pm\sqrt{\Theta(\theta)}, \\ \hat{\Sigma^2}\frac{d\phi}{d\lambda}&=\left(\hat{L}\csc^{2}\theta-a\hat{E}\right)+\frac{a}{\Delta}\left[\hat{E}(r^{2}+a^{2})-a\hat{L}\right].
\label{equationsformotions}
\end{align}
where
\begin{align}
R(r)&=\left[\hat{E}(r^{2}+a^{2})-a\hat{L}\right]^{2}-\Delta\left[\mathcal{J}+(\hat{L}-a\hat{E})^{2}\right],
\label{radialfunction}
\end{align}
\begin{align}
\Theta(\theta)&=\mathcal{J}+a^{2}\hat{E}^{2}\cos^{2}\theta-\hat{L}^{2}\cot^{2}\theta.
\label{Theatafunction}
\end{align}
where $\mathcal{J}$ is the Carter constant. Here, $R(r)$ and $\Theta(\theta)$ denote the radial and angular functions, respectively. The function $R(r)$ governs the radial motion of photons, whereas $\Theta(\theta)$ determines their motion in the polar direction. Together with the conserved quantities $\hat{E}$, $\hat{L}$, and the Carter constant $\mathcal{J}$, these functions completely describe the null geodesics around the Kerr-like black hole surrounded by CDM halo. Using these equations, one can determine the motion of photons around the rotating black hole.
\par
\begin{figure}[H]
\begin{center}
\subfigure[~$R_s=0.1,~\rho_c=0.05$]{\includegraphics[width=4.7cm,height=4.3cm]{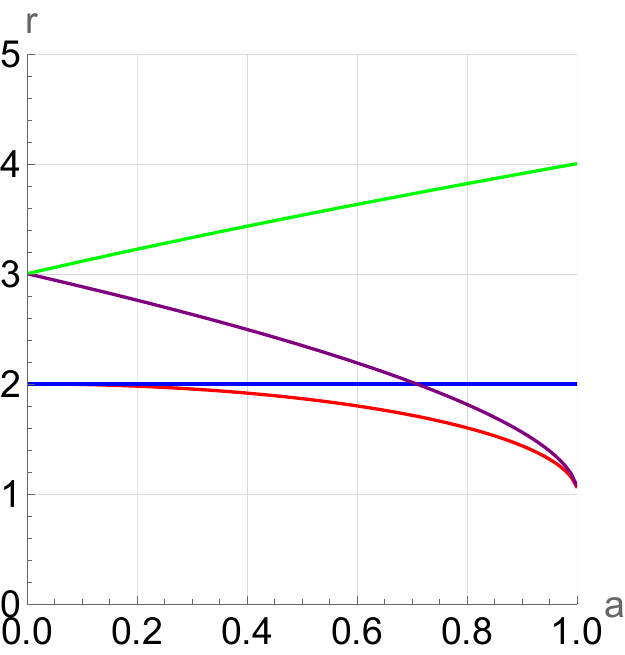}}
\subfigure[~$R_s=0.2,~\rho_c=0.05$]{\includegraphics[width=4.7cm,height=4.3cm]{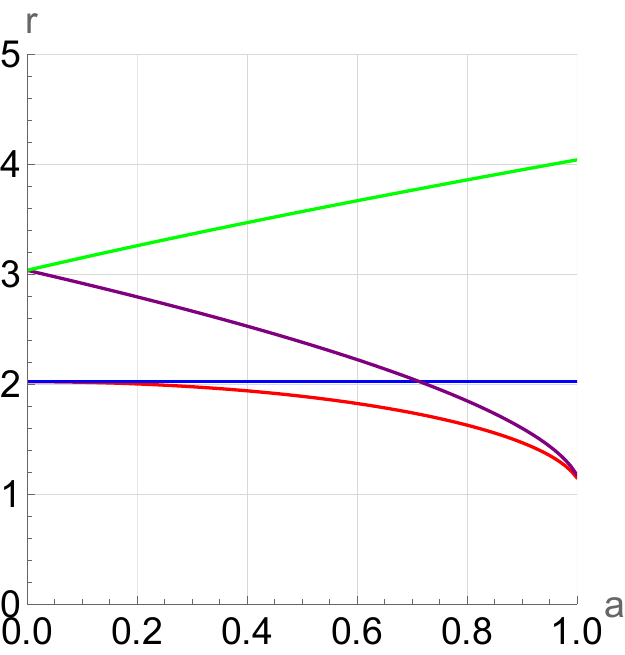}}
\subfigure[~$R_s=0.25,~\rho_c=0.05$]{\includegraphics[width=4.7cm,height=4.3cm]{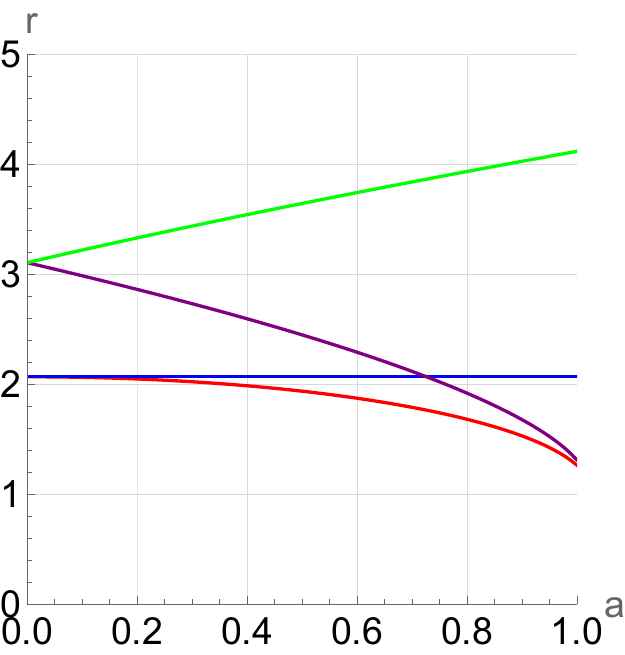}}
\subfigure[~$R_s=0.1,~\rho_c=0.25$]{\includegraphics[width=4.7cm,height=4.3cm]{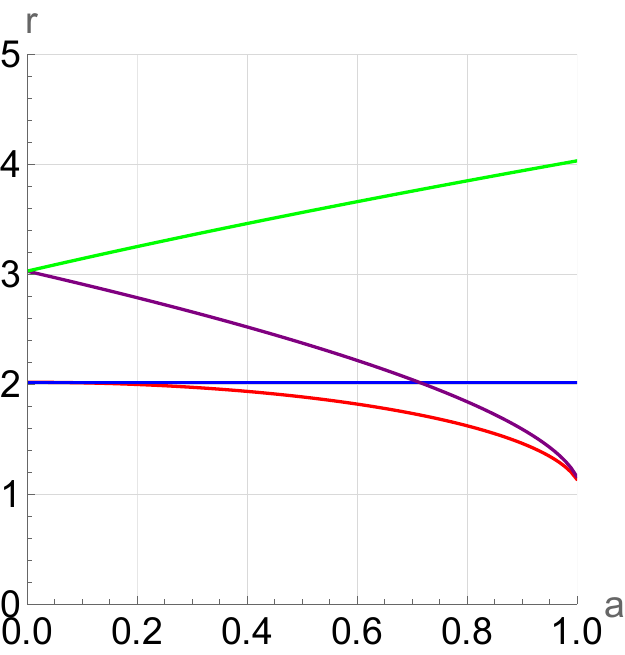}}
\subfigure[~$R_s=0.1,~\rho_c=0.45$]{\includegraphics[width=4.7cm,height=4.3cm]{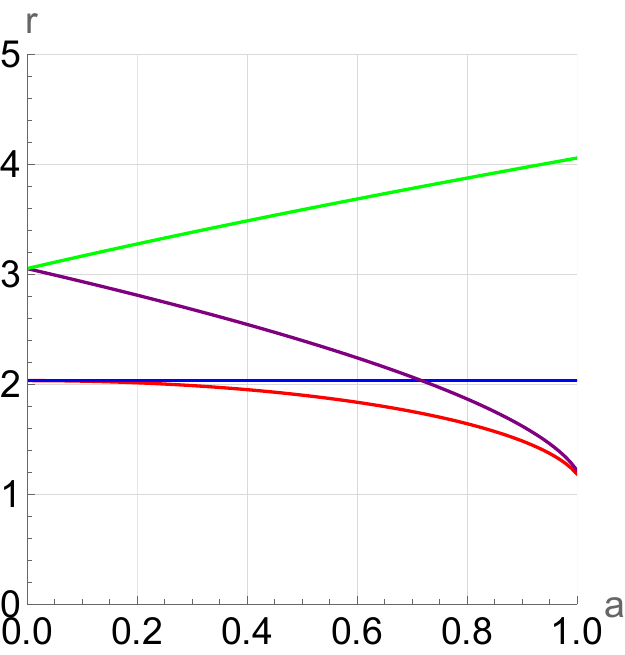}}
\subfigure[~$R_s=0.1,~\rho_c=0.65$]{\includegraphics[width=4.7cm,height=4.3cm]{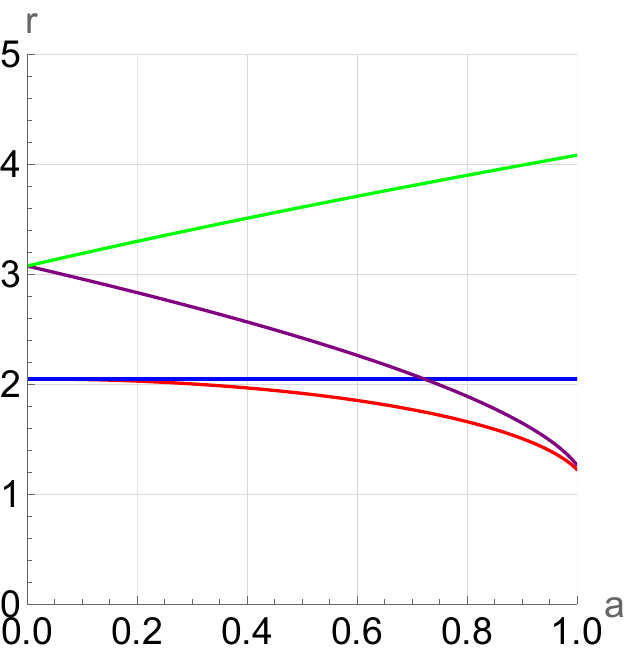}}
\caption{Plots showing the photon sphere radius, ergosphere boundary and event horizon with respect to $a$, under different values of $\rho_c$ and $R_s$. The green, purple, blue and red curves corresponds to the retrograde orbits, prograde orbits, the boundary of the ergosphere, and the event horizon, respectively.}
\label{fig1}
\end{center}
\end{figure}

 Considering the circular orbital motion of photons in equatorial plane ($\theta=\pi/2$), the radii of these circular orbits extend from infinity down to the photon sphere. The radius of the photon sphere satisfies the two conditions which are
\begin{equation}
R(r)=0, \quad R^{'}(r)=0.
\label{conditions}    
\end{equation}
These two conditions yield two solutions: one corresponding to prograde orbits, which co-rotate with the black hole, and the other corresponding to retrograde orbits, which counter-rotate relative to the black hole.
 In Fig. {\bf \ref{fig1}}, we plot the photon sphere radius, boundary of ergosphere, and event horizon $r$ as function of $a$. In Fig. {\bf \ref{fig1}}, the green, purple, blue and red curves corresponds to the retrograde orbits, prograde orbits, boundary of ergosphere, and event horizon, respectively. We can see that both critical density and scale radius reduce the maximum spin and the boundary of the ergosphere compared to the Kerr black hole. For the black hole surrounded by CDM halos, the maximum allowed spin is less than $1$, and the boundary of ergosphere is smaller than $2$. From the top row of Fig. {\bf\ref{fig1}}, we can see that, for the different values of $R_s$, the position of photon sphere radius slightly increase, but the boundary of ergosphere and event horizon remains the same. From the second row of Fig. {\bf\ref{fig1}}, also boundary of ergosphere and the event horizon remains similar, while the position of photon sphere slightly increases, under different values of $\rho_c$. Only prograde orbits are considered for energy extraction suggested by Fig. {\bf\ref{fig1}}. At the photon sphere radius, the energy and angular momentum of the particles approach infinity. Therefore, energy extraction actually occurs between the boundary of the ergosphere and the photon sphere radius.

\section{EXTRACTING ENERGY FROM ROTATING BLACK HOLE SURROUNDED BY CDM HALOS IN THE CIRCULAR ORBIT REGION}

\subsection{Magnetic Reconnection Process in Circular Orbits}

Here, we describe the formulas of Comisso-Asenjo \cite{comisso2021magnetic} which used for calculating the energy at infinity which are associated with accelerated and decelerated palsma in the spacetime (\ref{metric}). For calculating  the plasma energy density we use Zero Angular Momentum Observer (ZAMO) frame \cite{bardeen1972rotating}. The metric in ZAMO becomes Minkowskian metric which can be expressed as 
\begin{equation}
ds^{2}=-d\hat{t}^{\,2}+\sum_{i=1}^{3}(d\hat{x}^{\,i})^{2}=\eta_{\mu\nu}\,d\hat{x}^{\mu}d\hat{x}^{\nu},
\label{minkowskian}
\end{equation}
where we use 
\begin{equation}
d\hat{t}=\alpha\,dt,\qquad d\hat{x}^{\,i}=\sqrt{g_{ii}}\,dx^{i}-\alpha\beta^{i}dt,
\label{minkcomponents}
\end{equation}
along with 
\begin{equation}
\alpha=\sqrt{\left(-g_{tt}+\frac{g_{t\phi}^{2}}{g_{\phi\phi}}\right)},\qquad \beta^{\phi}=\frac{\omega^{\phi}}{\alpha}\sqrt{g_{\phi\phi}}.
\label{alphabeta}
\end{equation}
Here we define $\omega^\phi=-g_{t\phi}/g_{\phi\phi}$ as the angular velocity of the frame dragging due to the rotating regular spacetime. In Boyer-Lindquist (BL) coordinates, when we transform a contravariant vector $a^\mu$ into ZAMO frame which is denoted by $\hat{a}^{\mu}$, we obtain a following relation
\begin{equation}
\hat{a}^{0}=\alpha a^{0},\qquad \hat{a}^{i}=h_{i}a^{i}-\alpha\beta^{i}a^{0},
\label{contravariant}
\end{equation}
and we also transform a covariant vector $\hat{a}_{\mu}$ then we obtain  the following relation 
\begin{equation}
\hat{a}_{0}=\frac{a_0}{\alpha}+\sum_{i}\frac{\beta^{i}}{h_{i}}{a}_{i},\qquad \hat{a}_{i}=\frac{a_i}{h_{i}}.
\label{covariant}
\end{equation}
\par In BL coordinates the one-fluid approximation energy-momentum tensor is give in the form of 
\begin{equation}
T^{\mu\nu}=pg^{\mu\nu}+wU^{\mu}U^{\nu}+F^{\mu}{}_{\delta}F^{\nu\delta}-\frac{1}{4}g^{\mu\nu}F_{\rho\delta}F^{\rho\delta},
\label{energymomentumtensor}
\end{equation}
in the energy-momentum tensor $p$, $\omega$, $U^{\mu}$ and $F_{\mu\nu}$ are define as plasma pressure, enthalpy density, four-velocity, and electromagnetic field tensor, respectively. In this perspectives, one can obtain the energy-at-infinity density, which is defined as 
\begin{equation}
e^{\infty}=-\alpha g_{\mu 0}T^{\mu 0}=e^{\infty}_{\mathrm{hyd}}+e^{\infty}_{\mathrm{em}},
\label{einfinitydivision}
\end{equation}
where $e^{\infty}_{\mathrm{hyd}}$ is the hydrodynamic part of energy-at-infinity density, and $e^{\infty}_{\mathrm{em}}$ is the electromagnetic part of energy-at-infinity density, which are given as 
\begin{equation}
e^{\infty}_{\mathrm{hyd}}=\alpha\left(w\hat{\gamma}^{2}-p\right)+\alpha\beta^{\phi}w\hat{\gamma}^{2}\hat{v}^{\phi}, \\ \quad
e^{\infty}_{\mathrm{em}}=\frac{\alpha}{2}\left(\hat{B}^{2}+\hat{E}^{2}\right)+\left(\hat{\mathbf{B}}\times\hat{\mathbf{E}}\right)_{\phi},
\label{hydroandmeganeto}
\end{equation}
where $\hat{\gamma}=\hat{U}^0=1/\sqrt{1-\sum_{i=1}^{3}(d\hat{v}^i)^2}$ is the Lorentz factor, $\hat{B}^i=\epsilon^{ij}\hat{F}_{jk}/2$, is the component of magnetic field, and $\hat{E}^{i}=\hat{F}_{i0}$ is the component of electric field. In ZAMO frame $v^{\phi}$ is used for the azimuthal components of the plasma outfloe velocity. During the Comisso-Asenjo magnetic reconnection process, a considerable portion of magnetic energy is converted into kinetic energy of plasma, in the total energy one can ignore the contribution of $e^\infty_{\mathrm{em}}$, then we consider $e^{\infty} \approx e^{\infty}_{\mathrm{hyd}}$. Also we consider the plasma element is incompressible and adiabatic than 
\begin{equation}
e^{\infty}=\alpha\left(w\hat{\gamma}^{2}+\beta^{\phi}w\hat{\gamma}^{2}\hat{v}^{\phi}\right)+p\hat{\gamma}.
\label{neweinfinity}
\end{equation}
\par In order to investigate the localized magnetic reconnection process, we introduce the local rest frame as $\bar{x}^\mu=\left(\bar{x}^{0},\,\bar{x}^{1},\,\bar{x}^{2},\,\bar{x}^{3}\right)$, where the direction of $\bar{x}^1$ is parallel to radial direction, $\bar{x}^1=r$ and $\bar{x}^3$ is parallel to azimuthal direction, while $\bar{x}^3=\phi$. By the BL observer point of view, in the equatorial plane the local rest frame rotates with Keplerian angular velocity $\Omega_{K}$, which is given by 
\begin{equation}
\Omega_{K}=\frac{d\phi}{dt}=\frac{-g_{t\phi,r}+\sqrt{g_{t\phi,r}^{\,2}-(g_{tt,r})(g_{\phi\phi,r}})}{g_{\phi\phi,r}}.
\label{keplerianvelocity}
\end{equation}
The Keplerian velocity in the ZAMO frame according to the transformation relation of vectors between BL and ZAMO coordinates, is given as 
\begin{equation}
\hat{v}^K=\frac{\Omega_K}{\alpha}\sqrt{g_{\phi\phi}}-\frac{\omega^{\phi}}{\alpha}\sqrt{g_{\phi\phi}}.
\label{zamokeplerian}
\end{equation}
The symbol $v_{\mathrm{out}}$ is represent the outflow velocity in the local rest frame, than the outflow velocity in ZAMO frame is given as 
\begin{equation}    
v^{\phi}_{\pm}=\frac{\hat{v}^{K}\pm v_{\mathrm{out}}\cos(\xi)}{1\pm \hat{v}_{K}v_{\mathrm{out}}\cos(\xi)}
\label{outvelocity}
\end{equation}
where ($+$) show the co-rotating outflow velocity with black hole rotation, and ($-$) represent the counter-rotating outflow velocity with black hole rotation. They also used to represent the accelerated and decelerated parts of plasma respectively, Here $\xi$ is the plasma orientation angle. Using Eqs. (\ref{neweinfinity}) and (\ref{outvelocity}), one can obtain the energy-at-infinity density of the reconnection outflows in the form of 
\begin{equation}
e_{\pm}^{\infty}={\alpha\hat{\gamma}_{K}\left(\left(1+\hat{v}_{K}\beta^{\phi}\right)\gamma_{\mathrm{out}}w\pm\cos(\xi)\left(\hat{v}_{K}+\beta^{\phi}\right)\gamma_{\mathrm{out}}v_{\mathrm{out}}w-\frac{p}{\left(1\pm\cos(\xi)\hat{v}_{K}v_{\mathrm{out}}\right)\gamma_{\mathrm{out}}\hat{\gamma}_{K}^{2}}\right)}.
\label{energyatinfinitydensity}
\end{equation}
where $\gamma_{\mathrm{out}}=1/\sqrt{1-v_{\mathrm{out}}^2}$ and $\hat{\gamma}_{K}=1/\sqrt{1-\hat{v}_{K}}$. And $v_{\mathrm{out}}$ is related to the properties of plasma magnetization which can be expressed as 
\begin{equation}
v_{\mathrm{out}}=\sqrt{\frac{\sigma}{\sigma+1}}.
\label{magnetization}
\end{equation}
where $\sigma=B^2/\omega$ is the plasma magnetization upsttream of the reconnection layer, $B$ is the asymptotic macroscale magnetic field. Then the plasma energy-at-infinity density per enthalpy $\epsilon^{\infty}_{\pm}=e^{\infty}_{\pm}/\omega$ becomes \cite{comisso2021magnetic}
\begin{equation}
\epsilon_{\pm}^{\infty}=\alpha\hat{\gamma}_{K}\left[\left(1+\beta^{\phi}\hat{v}_{K}\right)(1+\sigma)^{1/2}\pm\cos(\xi)\left(\beta^{\phi}+\hat{v}_{K}\right)\sigma^{1/2}-\frac{1}{4}\frac{(1+\sigma)^{1/2}\mp\cos(\xi)\hat{v}_{K}\sigma^{1/2}}{\hat{\gamma}_{K}^{2}\left(1+\sigma-\cos^{2}(\xi)\hat{v}_{K}^{2}\sigma\right)}\right],
\label{epsilonenergy}
\end{equation}
here we assumed that $p=\omega/4$.
\par
\begin{figure}[H]
\begin{center}
\subfigure[~$a=0.98,~\rho_c=0.1,~R_s=0.2,~r=1.6$]{\includegraphics[width=8cm,height=5cm]{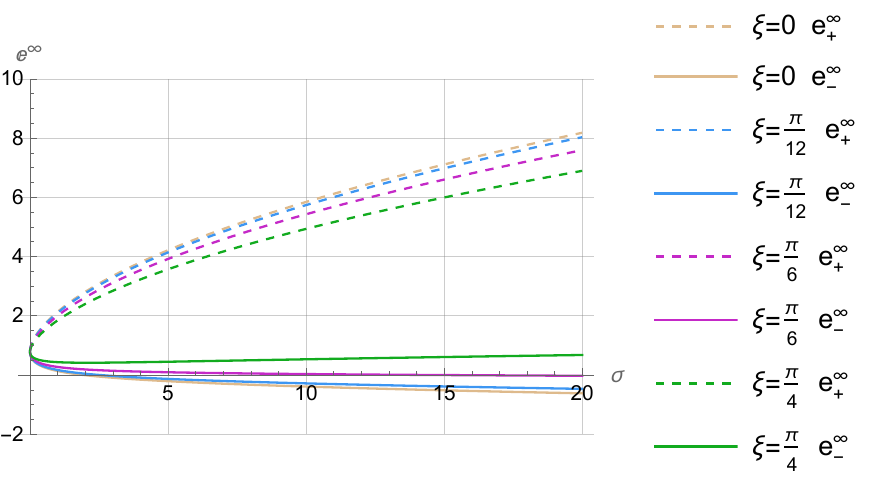}}
\subfigure[~$a=0.98,~\rho_c=0.1,~\xi=\pi/12,~r=1.9$]{\includegraphics[width=8cm,height=5cm]{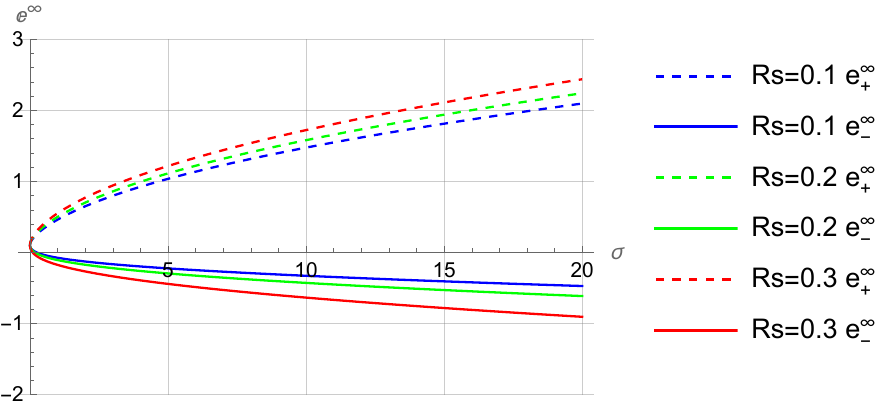}}
\caption{Plots showing the variation of $\epsilon^{\infty}_{+}$ and $\epsilon^{\infty}_{-}$ with respect to $\sigma$. The left panel illustrates the results for different values of $\xi$, whereas the right panel presents the results for different values of $R_s$.}\label{fig2}
\end{center}
\end{figure}

\begin{figure}[H]
\begin{center}
\subfigure[~$a=0.98,~R_s=0.1, ~\xi=\pi/12,~r=1.4$]{\includegraphics[width=8cm,height=5cm]{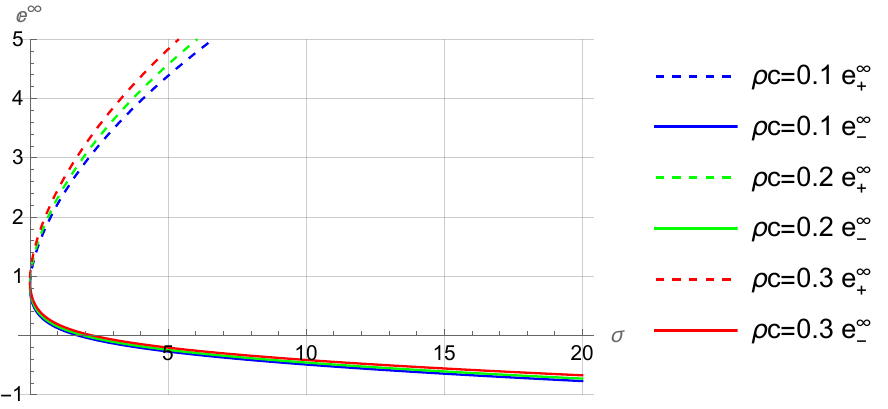}}
\subfigure[~$a=0.98,~R_s=0.1, ~\xi=\pi/12,~r=1.5$]{\includegraphics[width=8cm,height=5cm]{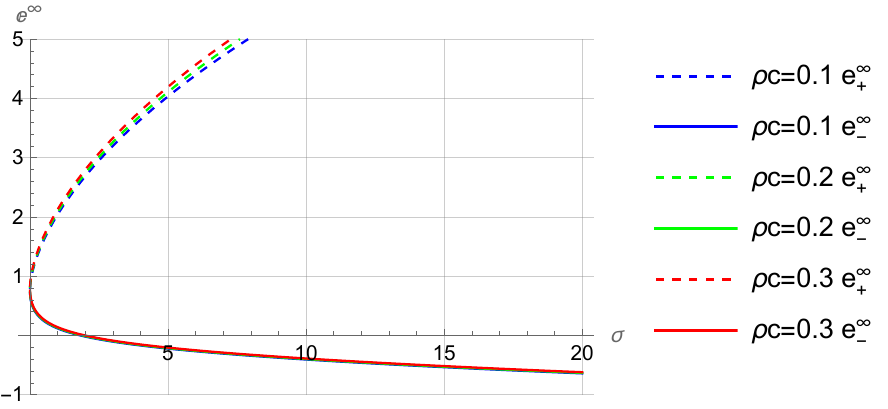}}
\caption{Plots showing the variation of $\epsilon^{\infty}_{+}$ and $\epsilon^{\infty}_{-}$ with respect to $\sigma$ under different values of $\rho_c$.}\label{fig3}
\end{center}
\end{figure}

 If the following conditions are satisfied, than the energy extraction mechanism can occurs,
\begin{equation}
\epsilon_{-}^{\infty}<0,\qquad \Delta\epsilon^{\infty}_{+}=\epsilon_{+}^{\infty}-\left(1-\frac{\Gamma}{4(\Gamma-1)}\right)>0,
\label{energyextcondition}
\end{equation}
here $\Gamma$ is a polytropic index, the value of $\Gamma$  for a relativistic hot plasma is taken as $4/3$. These conditions states that, the black hole energy can be extracted if the decelerated part of plasma possess negative energy measured at infinity, while accelerated part gain larger energy than its rest mass and thermal energies, in the same magnetic reconnection process. Now, we plot $\epsilon_{+}^{\infty}$ and $\epsilon_{-}^{\infty}$ in Figs. {\bf \ref{fig2}}, and {\bf \ref{fig3}}. Here $r$ is called dominant reconnection point, which is also known as X-point in \cite{comisso2021magnetic}. 
From Fig. {\bf\ref{fig2}}, it can be seen that $\epsilon^{\infty}_{+}$ remains positive throughout the entire range of the circular orbits, whereas $\epsilon^{\infty}_{-}$ is consistently negative. Since the magnetic reconnection energy extraction mechanism requires the ingoing plasma to possess negative energy, the condition $\epsilon_{-}^{\infty}<0$ must be satisfied for energy extraction to occur. We can also see that as $\sigma$ increases, $\epsilon^{\infty}_{-}$ decreases, while $\epsilon^{\infty}_{+}$ increases. From the left panel of {\bf\ref{fig2}}, it can be seen that, as the azimuthal angle increases, $\epsilon^{\infty}_{-}$ increases but $\epsilon^{\infty}_{+}$ decreases, which is similar to the general case. From the right panel of {\bf\ref{fig2}}, it can be also seen that, as $R_s$ increases, $\epsilon^{\infty}_{-}$ decreases and $\epsilon^{\infty}_{+}$ increases.
From Fig. {\bf\ref{fig3}} we observe that $\epsilon^{\infty}_{-}$ and $\epsilon^{\infty}_{+}$ do not vary monotonically with $\rho_c$, and depends on the location of the reconnection point. By comparing the both panels of {\bf\ref{fig3}}, it can be seen that, for the smaller values of $r$, the difference between dashed and solid curves is more clearer with the variations of $\rho_c$.

\subsection{Parameter Space for Energy Extraction Via Magnetic Reconnection in Circular Orbits}
\begin{figure}[H]
\begin{center}
\subfigure[~$\rho_c=0.05,R_s=0.1, ~\xi=\pi/12$]{\includegraphics[width=4.7cm,height=4.3cm]{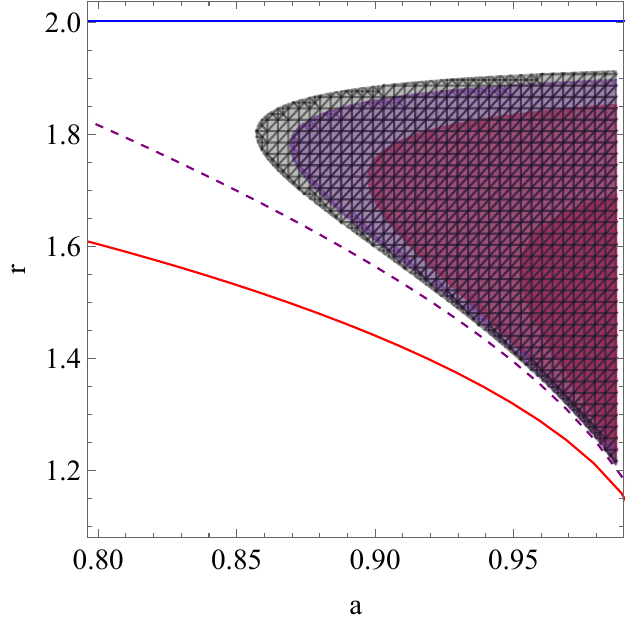}}
\subfigure[~$\rho_c=0.25, R_s=0.1, ~\xi=\pi/12$]{\includegraphics[width=4.7cm,height=4.3cm]{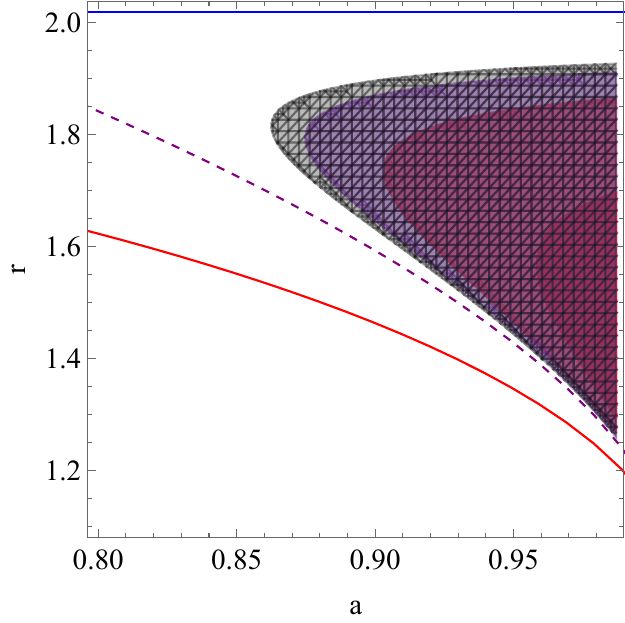}}
\subfigure[~$\rho_c=0.45, R_s=0.1, ~\xi=\pi/12$]{\includegraphics[width=4.7cm,height=4.3cm]{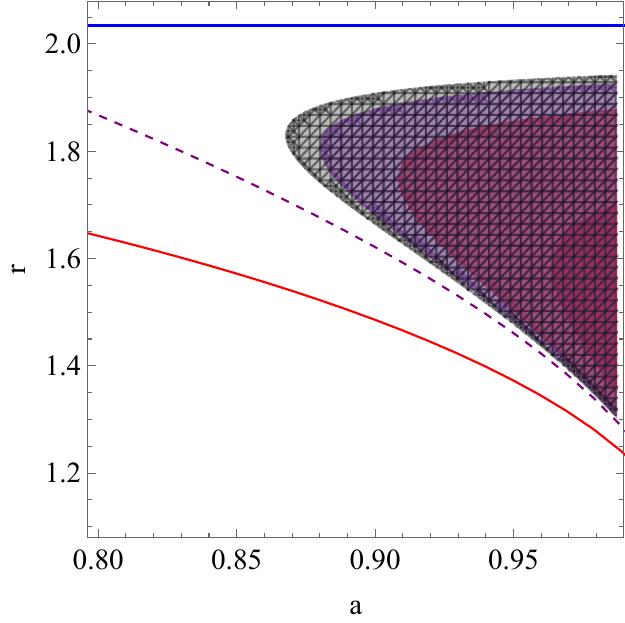}}
\caption{Plots showing the allowed energy extraction regions for different values of $\rho_c$.}
\label{fig4}
\end{center}
\end{figure}

\begin{figure}[H]
\begin{center}
\subfigure[~$R_s=0.1, \rho_c=0.05,  ~\xi=\pi/12$]{\includegraphics[width=4.7cm,height=4.3cm]{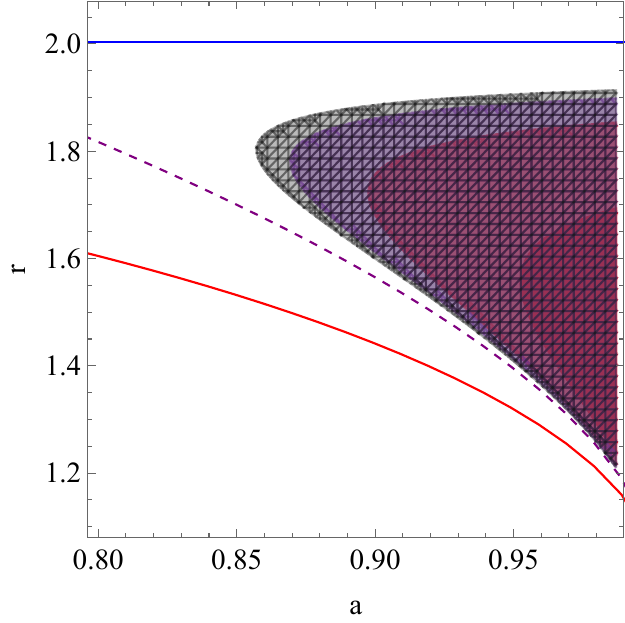}}
\subfigure[~$R_s=0.2, \rho_c=0.05,  ~\xi=\pi/12$]{\includegraphics[width=4.7cm,height=4.3cm]{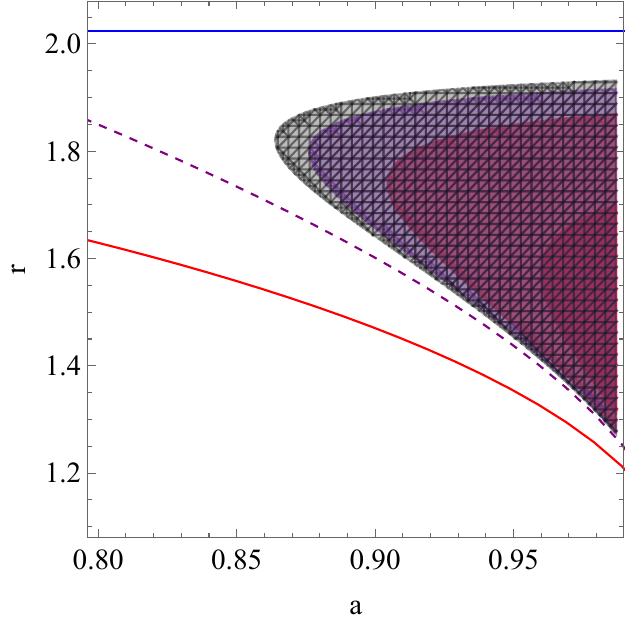}}
\subfigure[~$R_s=0.3, \rho_c=0.05,  ~\xi=\pi/12$]{\includegraphics[width=4.7cm,height=4.3cm]{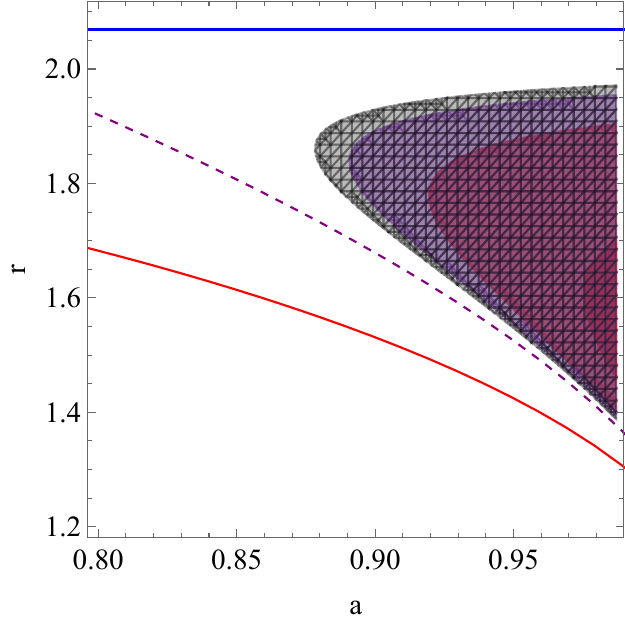}}
\caption{Plots showing the allowed energy extraction regions for different values of $R_s$.}
\label{fig5}
\end{center}
\end{figure}

\begin{figure}[H]
\begin{center}
\subfigure[~$ ~\xi=\pi/12, R_s=0.1, \rho_c=0.05$]{\includegraphics[width=4.7cm,height=4.3cm]{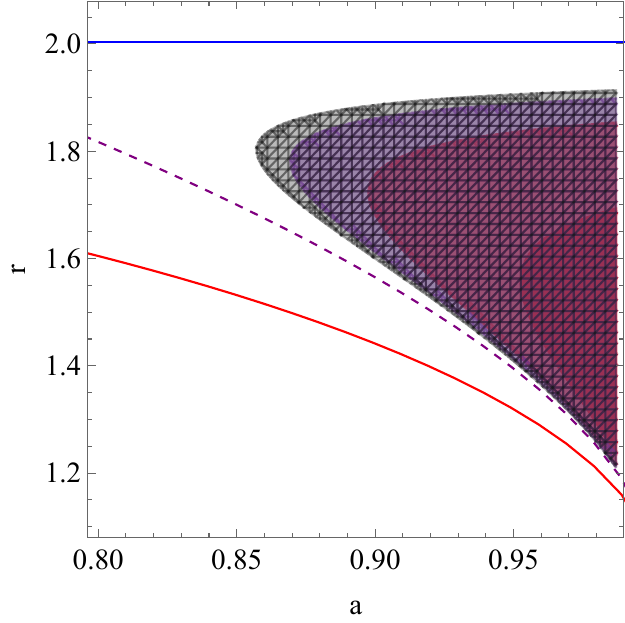}}
\subfigure[~$~\xi=\pi/6, R_s=0.1, \rho_c=0.05$]{\includegraphics[width=4.7cm,height=4.3cm]{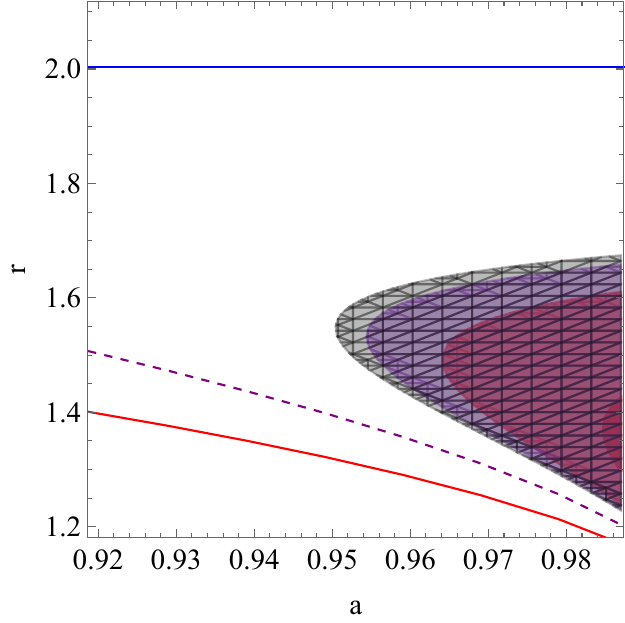}}
\subfigure[~$ ~\xi=0, R_s=0.1, \rho_c=0.05$]{\includegraphics[width=4.7cm,height=4.3cm]{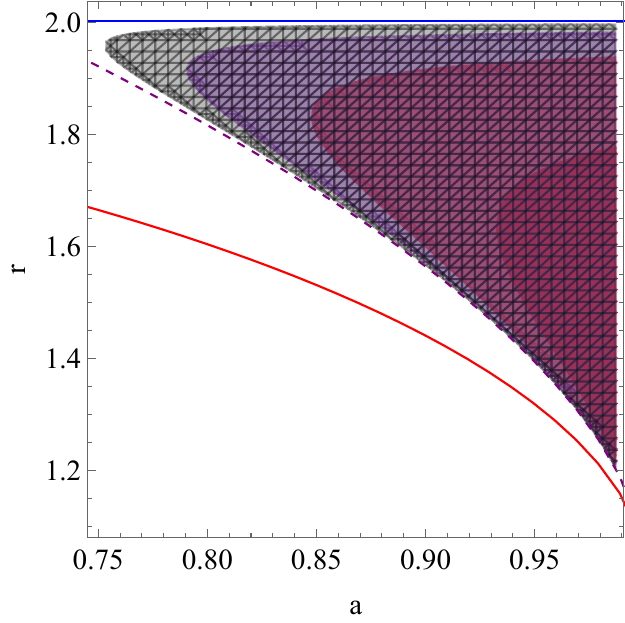}}
\caption{Plots showing the allowed energy extraction regions for different values of $\xi$.}
\label{fig6}
\end{center}
\end{figure}

\begin{figure}[H]
\centering 
\includegraphics[width=0.5\linewidth]{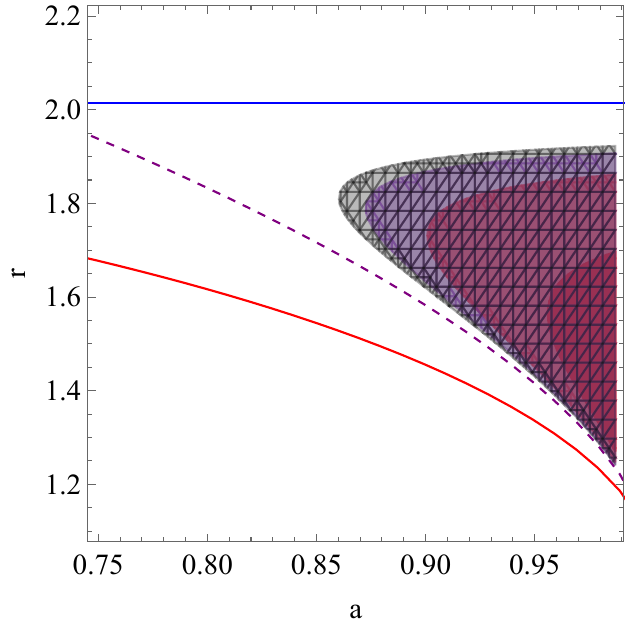}
\caption{Plot showing the allowed energy extraction region with a fixed value of $\rho_c=0.01$ and $R_s=0.3$.}
\label{fig7}
\end{figure}

In this section, we plot the allowed region where we can extract energy, such as $\epsilon^{\infty}_{-}<0$ in the $r-a$ plane in Figs. {\bf\ref{fig4}-\ref{fig7}}. Here, blue solid line represent the ergosphere, red solid line represents the event horizon, and purple dashed line represents the photon sphere radius, and from left to right $\sigma=100,~30,~10,~3$, respectively.
From Fig. {\bf\ref{fig4}}, it can be observed that as the value of $\sigma$ increases, the allowed region for energy extraction also expands. This indicates that larger values of $\sigma$ enhance the parameter space within which the magnetic reconnection mechanism can efficiently extract energy from the black hole, which is consistent with \cite{comisso2021magnetic}. Under the different values of $\rho_c$, the maximum allowed spin remain same, such as $0.987$. But the minimum allowed spin increases, as $\rho_c$ increases, from $0.86$ in Fig. {\bf\ref{fig4} (a)} to $0.88$ in {\bf\ref{fig4} (c)}. Similarly, as the ergosphere, the event horizon, and the photon sphere radius increase, the position of the reconnection layer also shifts outward.
From Fig. {\bf\ref{fig5}}, we notice that, as $R_s$ increases, the minimum allowed spin for energy extraction increases, from $0.86$ in {\bf\ref{fig5} (a)} to $0.90$ in {\bf\ref{fig5} (c)}, but the maximum allowed spin for energy extraction remain same, analogous to the effect of increasing $\rho_c$. In this case, the position of the reconnection layer shifts outward with the increasing of ergosphere, the event horizon, and the photon sphere radius. From our analysis, we conclude that, for circular orbits, both parameters, $\rho_c$ and $R_s$, favour energy extraction at lower black hole spin values.
From Fig. {\bf\ref{fig6}}, it can be observed that as the value of the angle $\xi$ decreases, the allowed region for energy extraction expands, while the minimum spin required for energy extraction decreases to approximately $a \simeq 0.76$. In addition, the position of the reconnection layer shifts outward. Since the influence of $\xi$ on the magnetic reconnection process is relatively weak, so we fixed its value at $\xi=\pi/12$ throughout the present analysis.
In Fig. {\bf\ref{fig7}}, there exists a minimum value of spin parameter required for magnetic reconnection to occur. We reveal that for the spacial values of parameters $\rho_c$ and $R_s$, there is low spin. Furthermore, our results show that the allowed spin for energy extraction lies between $0.86$ and $0.988$, demonstrating that the energy extraction process can take place at moderate black hole spin.

\subsection{Power and Efficiency of Energy Extraction in Circular Orbits}
After establishing the conditions required for energy extraction, we proceed to compare the power and efficiency of the magnetic reconnection process for circular orbits. The power extracted through magnetic reconnection is given by \cite{comisso2021magnetic}
\begin{equation}
P=-\epsilon^{\infty}_{-}\omega A_{\rm in}U_{\rm in}.
\label{powerext}
\end{equation}
 Two representative cases are generally considered: the collisionless reconnection regime, for which $U_{\rm in}\approx0.1$ \cite{comisso2016value}, and the collisional reconnection regime, for which $U_{\rm in}\approx0.01$ \cite{huang2010scaling,uzdensky2010fast}. In the present work, we consider only the first case, i.e., the collisionless reconnection regime with $U_{\rm in}\approx0.1$.
Furthermore, $A_{\rm in}$ denotes the cross-sectional area of the inflowing plasma and is approximated by
\begin{equation}
A_{\rm in}\sim \left(r_{\rm E}^{2}-r_{\rm ph}^{2}\right),
\label{crossarea}
\end{equation}
where $r_{\rm E}$ and $r_{\rm ph}$ represent the boundary of the ergosphere and the photon sphere radius, respectively. 
\par
\begin{figure}[H]
\begin{center}
\subfigure[~$a=0.98,~\xi=\pi/12,~\rho_c=0.05,R_s=0.1$]{\includegraphics[width=8cm,height=5cm]{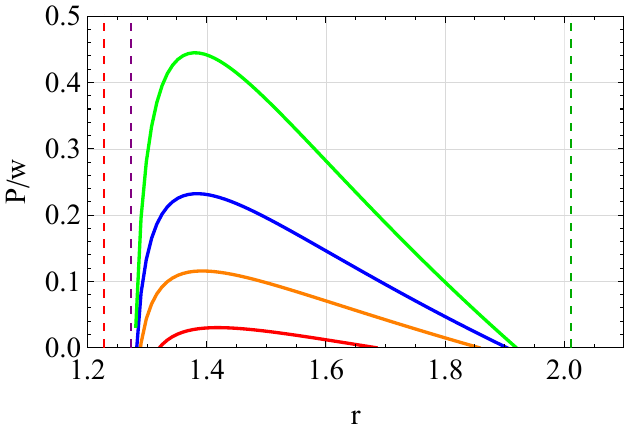}}
\subfigure[~$a=0.96,~\xi=\pi/12,~\rho_c=0.15,R_s=0.1$]{\includegraphics[width=8cm,height=5cm]{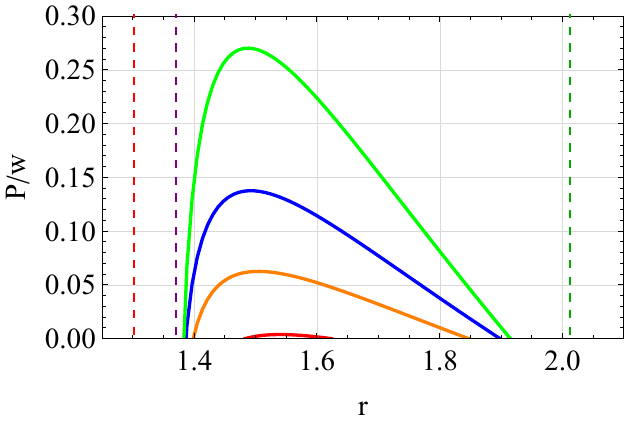}}
\caption{Plots showing the energy extraction power for different values of $\sigma$.}
\label{fig8}
\end{center}
\end{figure}

\begin{figure}[H]
\begin{center}
\subfigure[~$a=0.94,~\xi=\pi/12,~\rho_c=0.05, R_s=0.17$]{\includegraphics[width=8cm,height=5cm]{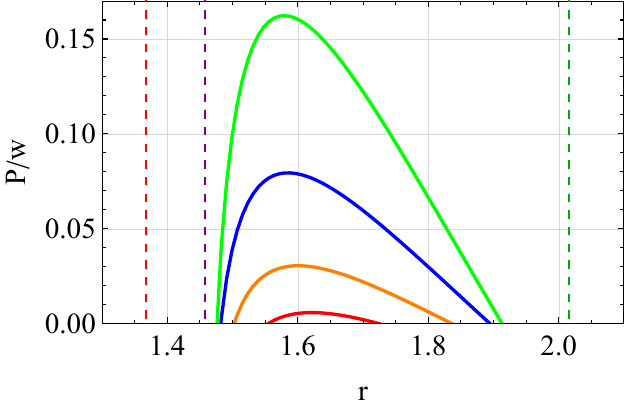}}
\subfigure[~$a=0.92,~\xi=\pi/12,~\rho_c=0.05,R_s=0.26$]{\includegraphics[width=8cm,height=5cm]{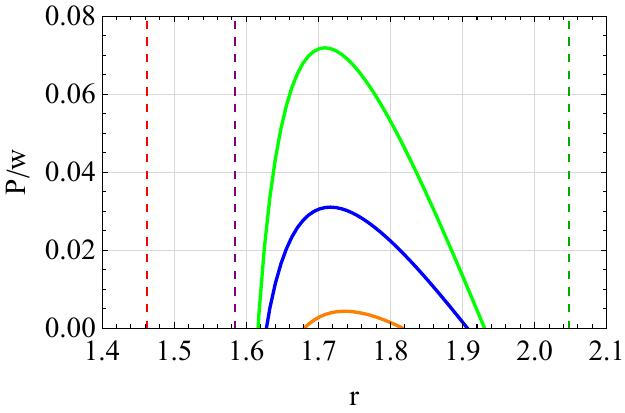}}
\caption{Plots showing the energy extraction power for different values of $\sigma$.}
\label{fig9}
\end{center}
\end{figure}

 Now we plot the energy extraction power per enthalpy density $p/\omega$ with respect to $r$ in Figs. {\bf\ref{fig8}} and {\bf\ref{fig9}}. Where red dashed line represent the event horizon, purple dashed line represent the photon sphere and green dashed line represent the boundary of ergosphere. Further, the red, orange, blue, and green solid curves represents the $\sigma=3,~10,~30,~100$, rspectively.
From Fig. {\bf\ref{fig8}}, it can be observed that the power starts from outside the photon sphere, which is consistent with the properties of circular orbits. As the orbital radius increases, the power initially increases, reaches a maximum, and then gradually decreases. Moreover, the extracted power increases with increasing values of $\sigma$. A comparison of the two panels in Fig. {\bf\ref{fig8}} further shows that, for the same value of $\sigma$, the extracted power decreases as the value of $\rho_c$ increases. It is also evident that the spin parameter $a$ is not the same in the two panels. This is because the allowed black hole spin vary under different value of $\rho_c$, resulting in different values of $a$ for each panel. This behavior is consistent with that reported in Ref. \cite{l2023energy}, where the spin parameter also varies with different values of the corresponding $h_0$ parameter. A comparison between Figs. {\bf\ref{fig8}(a)} and {\bf\ref{fig9}(a)} shows that the magnetic reconnection power decreases with increasing values of $R_s$, primarily due to the lower black hole spin required for energy extraction. It can also be observed that the spin parameter $a$ is not identical in the two figures, since the allowed black hole spin varies for different values of $R_s$. For selected values of $R_s$ and $\rho_c$, the energy extraction power is further illustrated in Fig. {\bf\ref{fig9} (b)}. It is evident that the extracted power is lower than that obtained in the previous cases, indicating that these parameter choices reduce the efficiency of the magnetic reconnection energy extraction mechanism.
\par
\begin{figure}[H]
\begin{center}
\subfigure[~$\rho_c=0.05,R_s=0.1$]{\includegraphics[width=8cm,height=5cm]{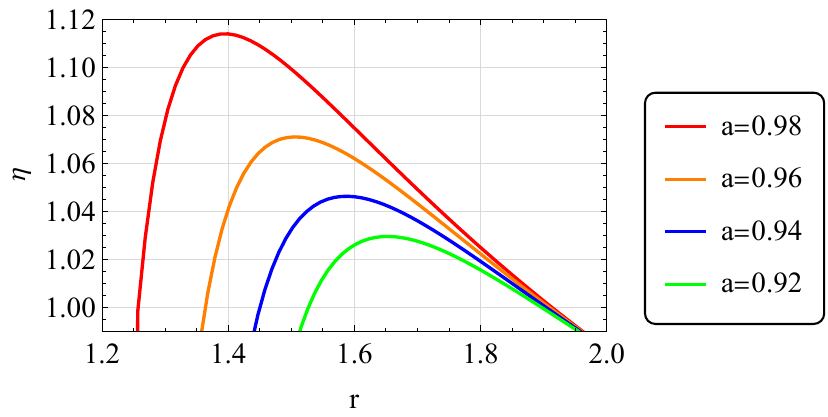}}
\subfigure[~$\rho_c=0.05,R_s=0.25$]{\includegraphics[width=8cm,height=5cm]{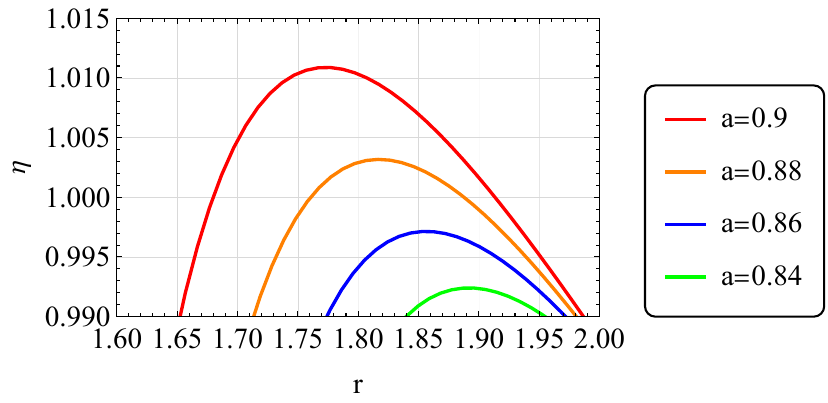}}
\caption{Plots showing the energy extraction efficiency for different values of $a$.}
\label{fig10}
\end{center}
\end{figure}

\begin{figure}[H]
\begin{center}
\subfigure[~$\rho_c=0.25,R_s=0.1$]{\includegraphics[width=8cm,height=5cm]{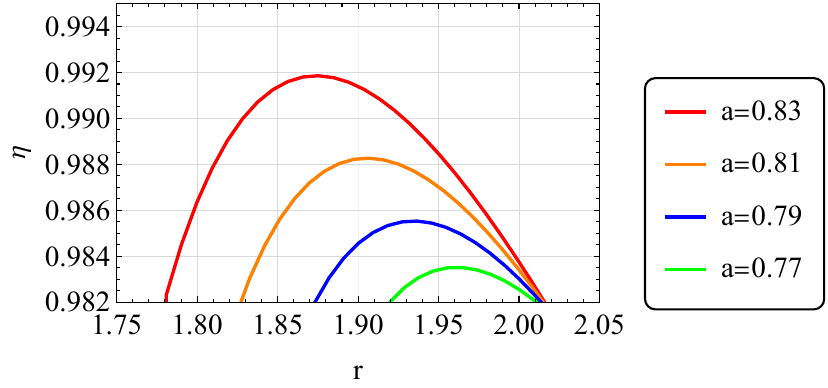}}
\subfigure[~$\rho_c=0.3,R_s=0.12$]{\includegraphics[width=8cm,height=5cm]{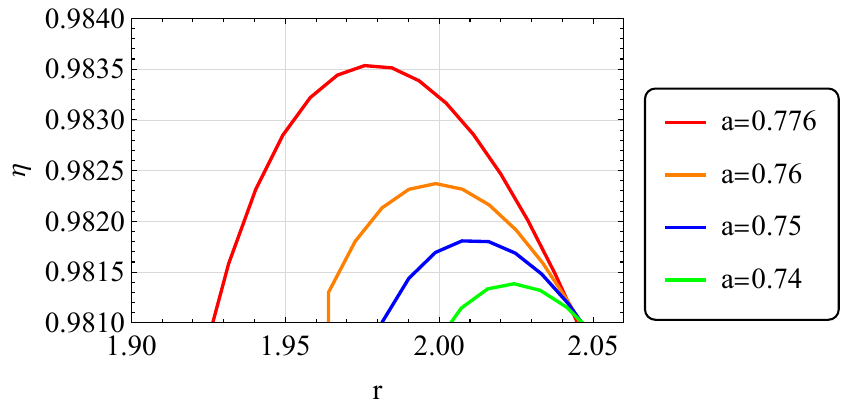}}
\caption{Plots showing the energy extraction efficiency for different
values of $a$.}
\label{fig11}
\end{center}
\end{figure}

 We next discuss the efficiency of the magnetic reconnection energy extraction mechanism, which is defined as \cite{comisso2021magnetic}
\begin{equation}
\eta=\frac{\epsilon_{+}^{\infty}}{\epsilon_{+}^{\infty}+\epsilon_{-}^{\infty}}.
\label{efficiency}
\end{equation}
As discussed previously, the energy extraction process requires the conditions $\epsilon_{-}^{\infty}<0$ and $\epsilon_{+}^{\infty}>0$ to be satisfied. Consequently, whenever the magnetic reconnection mechanism operates successfully, the corresponding energy extraction efficiency must satisfy $\eta>1$. The efficiency of the energy extraction process is plotted as a function of $r$ in Figs. {\bf\ref{fig10}} and {\bf\ref{fig11}}. Throughout this analysis, we fix the parameters at $\xi=\pi/12$ and $\sigma=100$.
Figure {\bf\ref{fig10} (a)} illustrates the energy extraction efficiency for different values of the black hole spin parameter $a$ at a relatively small value of $\rho_c$. Similar to the behaviour of the extracted power, the efficiency initially increases, reaches a maximum, and then gradually decreases. Moreover, the efficiency increases with increasing values of the spin parameter $a$. The corresponding results for a larger value of $\rho_c$ are presented in Fig. {\bf\ref{fig11} (a)}. A comparison between Figs. {\bf\ref{fig10} (a)} and {\bf\ref{fig11} (a)} shows that the energy extraction efficiency decreases as $\rho_c$ increases. The observed trend is therefore consistent with that obtained for the extracted power. It should also be noted that the values of the spin parameter $a$ are not identical in Figs. {\bf\ref{fig10} (a)} and {\bf\ref{fig11} (a)}, since the minimum allowed black hole spin for energy extraction varies with different values of $\rho_c$. Next, we investigate the effect of $R_s$ on the energy extraction efficiency while keeping the $\rho_c$ is fixed. A comparison between Figs. {\bf\ref{fig10} (a)} and {\bf\ref{fig10} (b)} shows that the energy extraction efficiency decreases with increasing values of $R_s$. The observed behavior is consistent with the trend obtained for the extracted power. Furthermore, the energy extraction efficiency is plotted for selected values of $R_s$ and $\rho_c$ in Fig. {\bf\ref{fig11} (b)}. Comparing Figs. {\bf\ref{fig10} (a)} and {\bf\ref{fig11} (b)}, it is evident that the efficiency is significantly reduced for these parameter values. This behavior is closely follows the same trend, as observed for the energy extraction power.
Finally, we compare the power extracted through the magnetic reconnection mechanism with that of the well-known Blandford–Znajek mechanism. The energy extraction power associated with the Blandford–Znajek process is given by \cite{tchekhovskoy2010black}
\begin{equation}
P_{\rm BZ}=\frac{\kappa}{16\pi}\Phi_H^2\left(\Omega_H^2+c_1\Omega_H^4+c_2\Omega_H^6+\mathcal{O}(\Omega_H^8)\right),
\label{bzpowerforratio}
\end{equation}
where $\kappa$, $c_1$ and $c_2$ are numerical constants , which have the values as $0.05, 1.38$, and $-9.2$, respectively. $\Phi_{H}$ is the magnetic flux of the black hole event horizon, which is given as 
\begin{equation}
\Phi_{H}=\frac{1}{2}\int\int\left|B^{r}\right|\sqrt{g_{\theta\theta}g_{\phi\phi}}\,d\theta\,d\phi=2\pi\left(r_{+}^{2}+a^{2}\right)B_{0}\sin(\xi).
\label{magneticflux}
\end{equation}
where $r_{+}$ is the radius of event horizon, $B_{0}=(\omega\sigma)^{1/2}$, and $\Omega_{H}$ is the event horizon angular velocity, which is given as 
\begin{equation}
\Omega_{H}=\frac{-g_{t\phi}}{g_{\phi\phi}}|_{r=r_{+}}=\frac{a}{r^2_{+}+a^2}.
\label{angularvelocity}
\end{equation}
Simplifying the Eqs. (\ref{powerext}), (\ref{bzpowerforratio}), (\ref{magneticflux}) and (\ref{angularvelocity}), gives the power ratio as in the form of 
\begin{equation}
P_{\rm BZ}=\frac{-4\epsilon^{\infty}_{-}A_{in}U_{in}}{\kappa\pi\sigma(\Omega^2_{H}+c_1\Omega^4_{H}+c_2\Omega^6_{H})\sin^2(\xi)(r_{+}^2+a^2)^2}.
\label{powerratio}
\end{equation}
It can be seen from Eq. (\ref{powerratio}) that, as long as the azimuthal angle $\xi$ remains sufficiently small, the power ratio can always be greater than unity. This implies that the power extracted through the magnetic reconnection mechanism exceeds that of the Blandford-Znajek mechanism. Therefore, throughout the present analysis, we fix the azimuthal angle at $\xi=\pi/12$. So, we take parameters for high spin, for example $\rho_c=0.05,~R_s=0.2,~\sigma=7,~r=1.4,~a=0.98$. We obtain a power ratio of $4.08458>1$, which confirms that the magnetic reconnection power exceeds the Blandford-Zanjek power. Now we pick a set of parameters for low spin, for example, $\rho_c=0.45,~R_s=0.1,~\sigma=3,~r=2.2,~a=0.70$. We obtain a power ratio $1.32881$. For this set of parameters, even though the absolute power and efficiency are reduced but the magnetic reconnection power still exceeds the Blandford-Zanjek power. This is thanks to the factors $\rho_c$ and $R_s$, which lower the spin threshold and reflect the high efficiency of magnetic reconnection.

\section{EXTRACTING ENERGY FROM ROTATING BLACK HOLE SURROUNDED BY CDM HALOS IN THE PLUNGING REGION}

\subsection{Magnetic Reconnection Process in the Plunging Region}
\begin{figure}[H]
\begin{center}
\subfigure[~$\rho_c=0.05, R_s=0.1, ~\xi=\pi/12$]{\includegraphics[width=4.7cm,height=4.3cm]{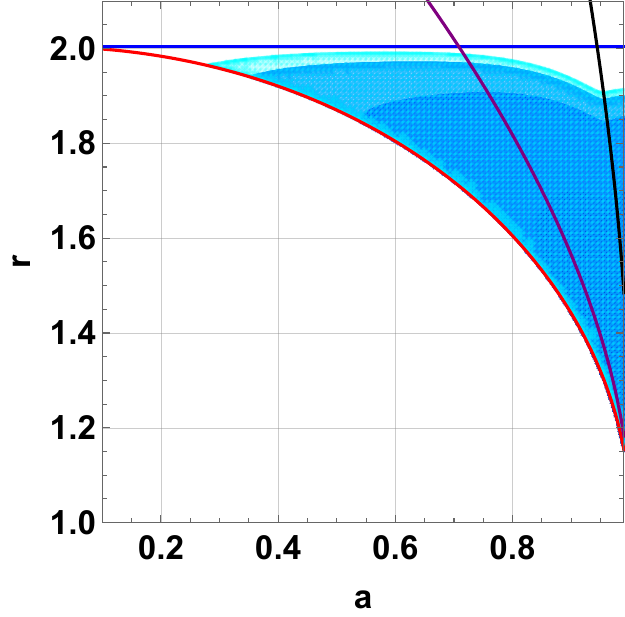}}
\subfigure[~$\rho_c=0.25, R_s=0.1, ~\xi=\pi/12$]{\includegraphics[width=4.7cm,height=4.3cm]{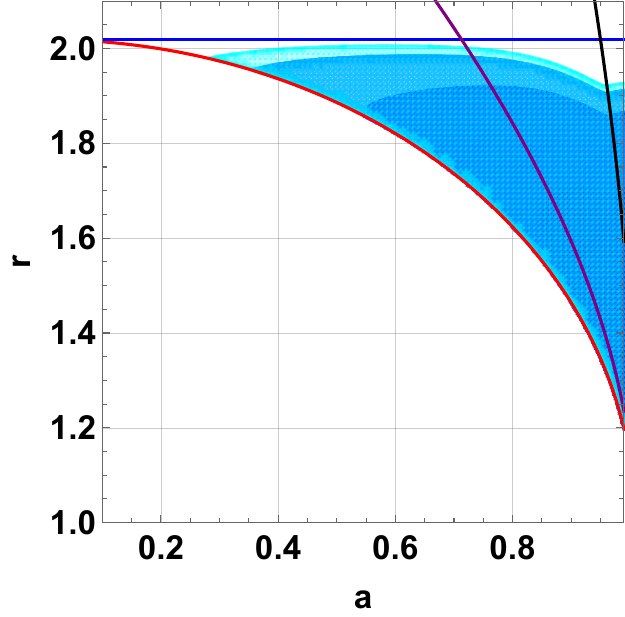}}
\subfigure[~$\rho_c=0.45, R_s=0.1, ~\xi=\pi/12$]{\includegraphics[width=4.7cm,height=4.3cm]{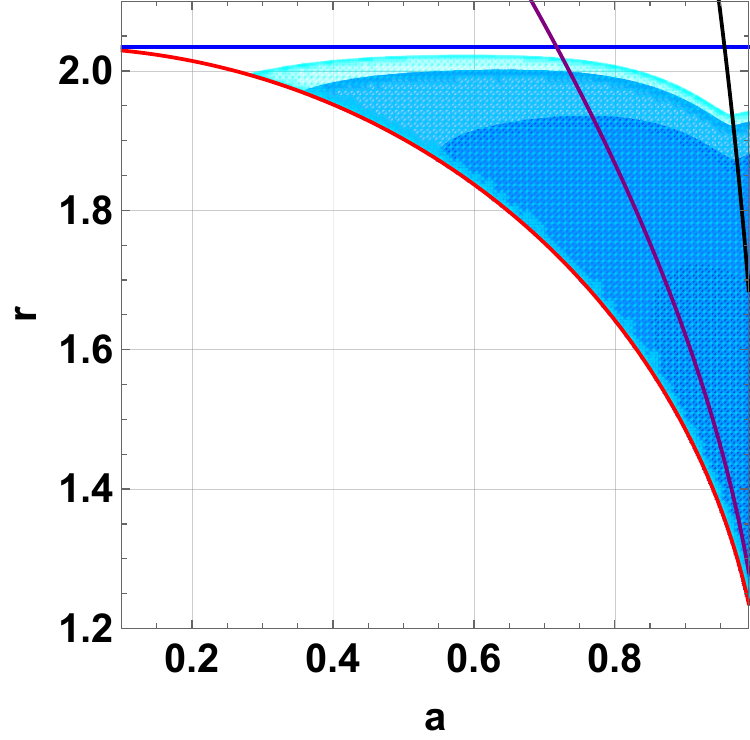}}
\caption{Plots showing the allowed energy extraction regions for different values of $\rho_c$ in the plunging region.}
\label{fig12}
\end{center}
\end{figure}

\begin{figure}[H]
\begin{center}
\subfigure[~$ R_s=0.1, \rho_c=0.05, ~\xi=\pi/12$]{\includegraphics[width=4.7cm,height=4.3cm]{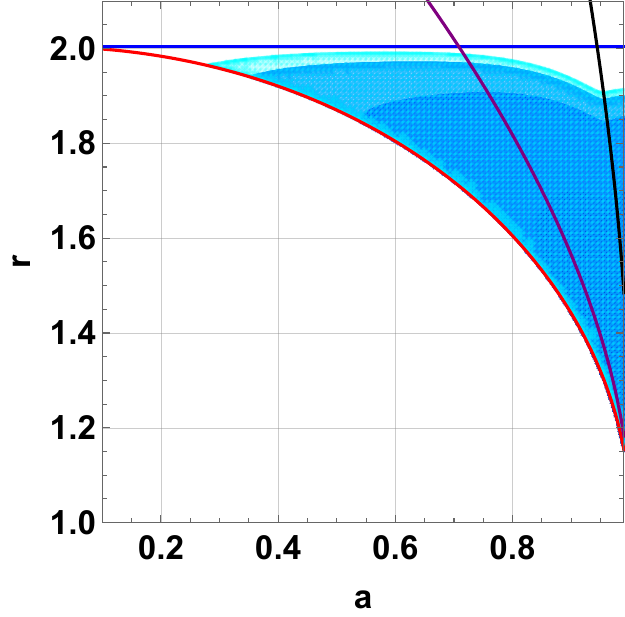}}
\subfigure[~$R_s=0.2, \rho_c=0.05,  ~\xi=\pi/12$]{\includegraphics[width=4.7cm,height=4.3cm]{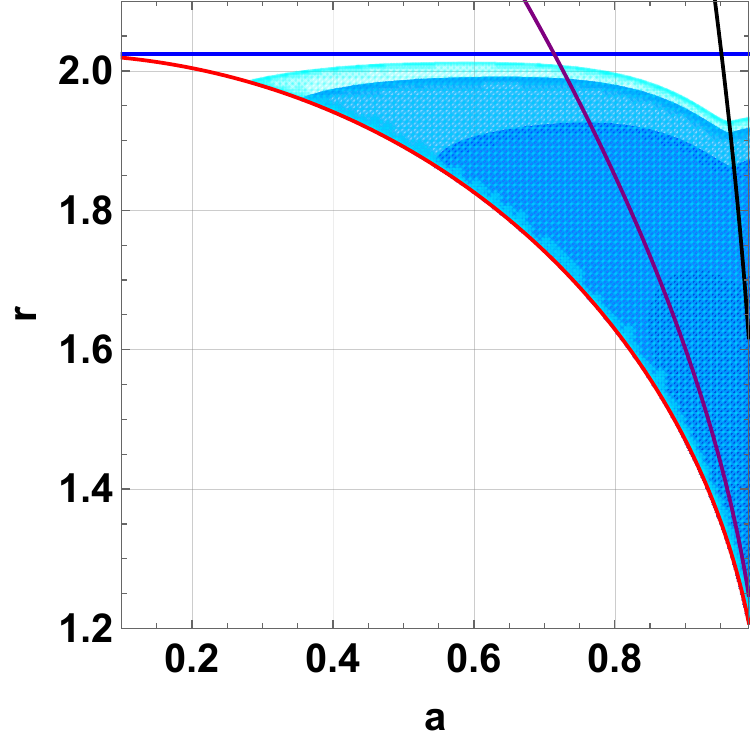}}
\subfigure[~$ R_s=0.3, \rho_c=0.05, ~\xi=\pi/12$]{\includegraphics[width=4.7cm,height=4.3cm]{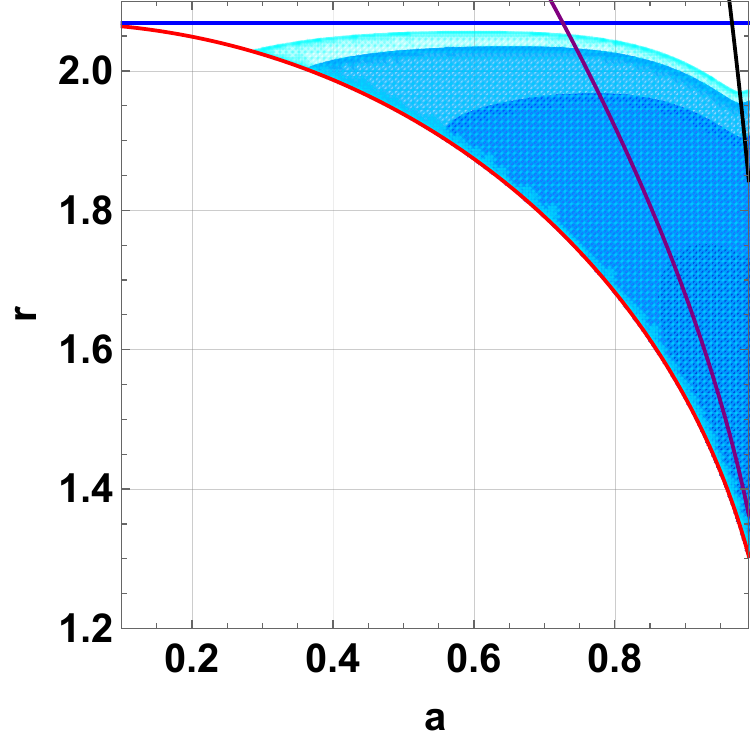}}
\caption{Plots showing the allowed energy extraction regions for different values of $R_s$ in the plunging region.}
\label{fig13}
\end{center}
\end{figure}

\begin{figure}[H]
\begin{center}
\subfigure[~$~\xi=\pi/12, \rho_c=0.05, R_s=0.1$]{\includegraphics[width=4.7cm,height=4.3cm]{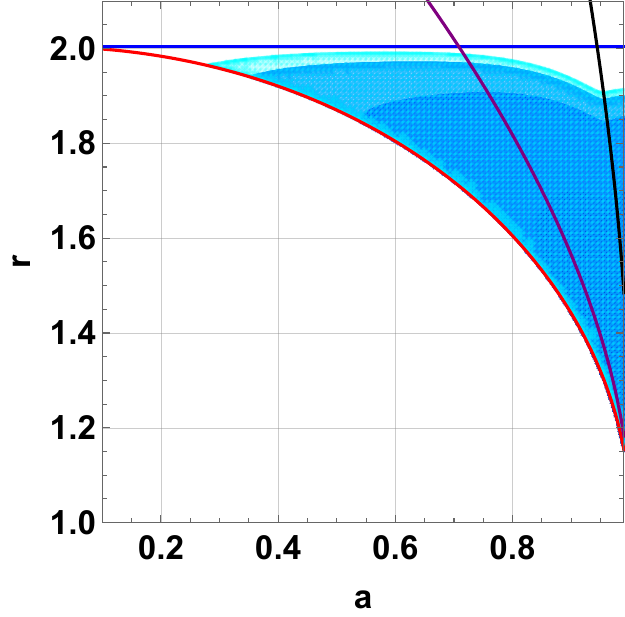}}
\subfigure[~$ ~\xi=\pi/6, \rho_c=0.05, R_s=0.1$]{\includegraphics[width=4.7cm,height=4.3cm]{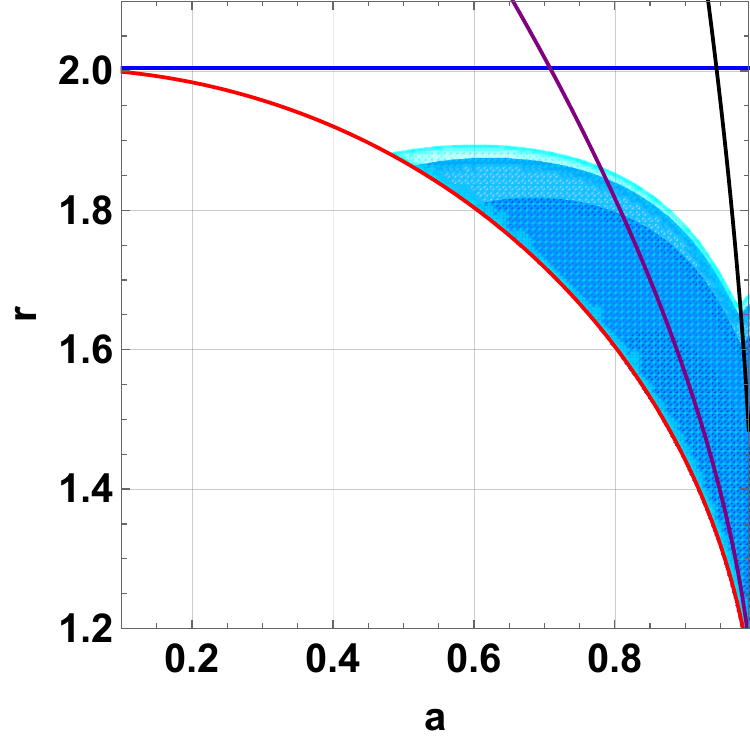}}
\subfigure[~$ ~\xi=0, \rho_c=0.05, R_s=0.1$]{\includegraphics[width=4.7cm,height=4.3cm]{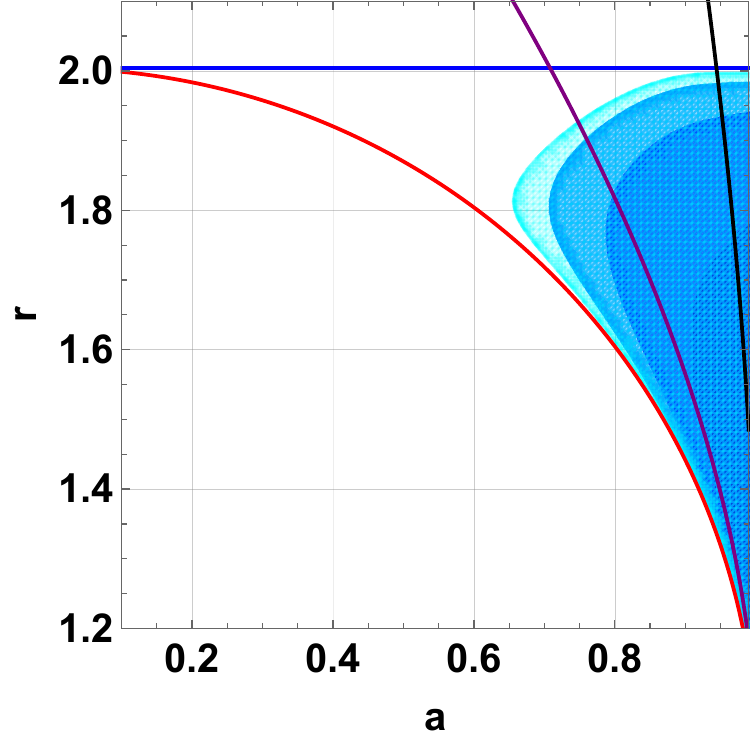}}
\caption{Plots showing the allowed energy extraction regions for different values of $\xi$ in the plunging region.}
\label{fig14}
\end{center}
\end{figure}

\begin{figure}[H]
\centering 
\includegraphics[width=0.5\linewidth]{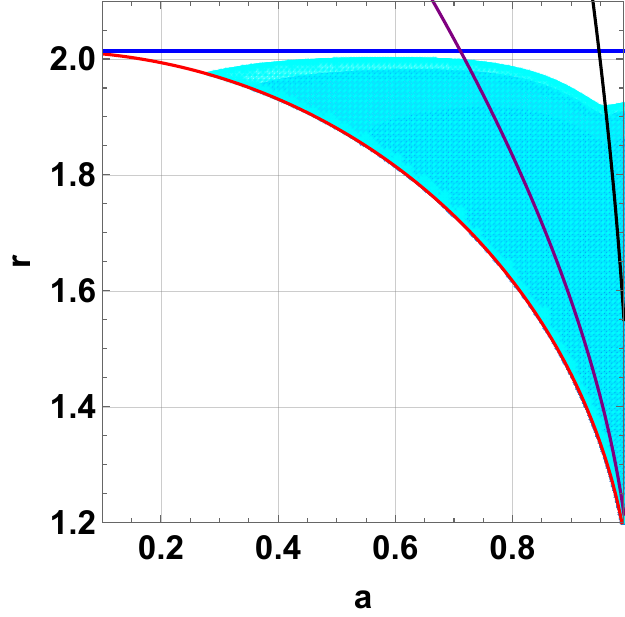}
\caption{ Allowed energy extraction region for $\rho_c=0.01,~R_s=0.32,~\xi=\pi/12$ in the plunging region.
}
\label{fig15}
\end{figure}

In the previous sections, we investigated the magnetic reconnection process by considering plasma moving in circular orbits outside the  photon sphere radius. We now turn to a different physical scenario \cite{shen2024energy} in which the plasma initially follows a stable circular orbit outside the innermost stable circular orbit (ISCO). As the plasma gradually loses angular momentum, it begins to move inward and eventually crosses the ISCO. Since circular orbits inside the ISCO are dynamically unstable, the plasma can no longer maintain stable circular motion. Instead, it acquires a significant radial velocity and starts plunging toward the black hole. The region defined by $r<r_{I}$, where $r_{I}$ denotes the radius of the ISCO and is larger than the photon sphere radius, is commonly referred to as the plunging region \cite{wilkins2020venturing}. In the plunging region, the plasma acquires a significant radial velocity component. Consequently, the Keplerian velocity expression given in Eq. (\ref{epsilonenergy}), which contains only the azimuthal component of the plasma motion, is no longer applicable. We still use the convenient ZAMO frame. The relationship between the four-velocity in the ZAMO frame and the BL frame is given as
\begin{equation}
U^{\mu}=\hat{\gamma}_{s}\begin{pmatrix}1 \\\hat{v}_{s}^{(r)} \\0 \\\hat{v}_{s}^{(\phi)}\end{pmatrix}=\begin{pmatrix}\dfrac{\hat{E}-\omega^{\phi}\hat{L}}{\alpha} \\\sqrt{g_{rr}}\,U^{r} \\0 \\ \dfrac{\hat{L}}{\sqrt{g_{\phi\phi}}} \end{pmatrix},
\label{BLcordinates}
\end{equation}
where
\begin{equation}
U^{r}=\frac{dr}{d\lambda},\qquad\left(U^{r}\right)^{2}=\left(\frac{dr}{d\lambda}\right)^{2}.
\label{urequation}
\end{equation}
In the plunging region, the energy $E$ and angular momentum $L$ are conserved quantities and their values at ISCO are replaced by 
\begin{equation}
\hat{L}_{I}=\hat{L}(r_{I}),\qquad \hat{E}_{I}=\hat{E}(r_{I}).
\label{replacement}
\end{equation} 
For ISCO, the conditions are 
\begin{equation}
R(r)=0,\qquad R'(r)=0,\qquad R''(r)=0.
\label{conditionforISCO}
\end{equation}
Substituting (\ref{replacement}) into (\ref{radialfunction}),  we get
\begin{align}
U^r
= -\frac{1}{\hat{\Sigma}^2}\sqrt{[(r^{2}+a^{2})\hat{E}_{I} - a\hat{L}_{I}]^2 - \Delta_r\left[(\hat{L}_{I}-a\hat{E}_{I})^2 + \mathcal{J}\right]}.
\label{uraftersubstution}
\end{align}
The negative sign in front of the above equation indicates the inward motion. Substituting (\ref{uraftersubstution}) into (\ref{urequation}) gives
$\hat{v}_{s}^{(r)}$,
$\hat{v}_{s}^{(\phi)}$,
and
\begin{equation}
\hat{v}_{s}=\sqrt{\left(\hat{v}_{s}^{(r)}\right)^{2}+\left(\hat{v}_{s}^{(\phi)}\right)^{2}},
\label{velocityv}
\end{equation}
where $\hat{\gamma}_{s}$ is the Lorentz factor of $\hat{v}_{s}$,
$r_I$ satisfies (\ref{conditionforISCO}).
In this case, $\epsilon_{\pm}^{\infty}$ becomes as \cite{comisso2021magnetic}
\begin{equation}
\begin{aligned}
\epsilon_{\pm}^{\infty}=&\,\alpha \hat{\gamma}_{s}\gamma_{\rm out}\Bigg[\left(1+\beta^{\phi}\hat{v}_{s}^{(\phi)}\right)\pm v_{\rm out}\left(\hat{v}_{s}+\beta^{\phi}\frac{\hat{v}_{s}^{(\phi)}}{\hat{v}_{s}}\right)\cos\xi\mp v_{\rm out}\beta^{\phi}\frac{\hat{v}_{s}^{(r)}}{\hat{\gamma}_{s}\hat{v}_{s}}\sin\xi\Bigg]\\&-\alpha\left[4\hat{\gamma}_{s}\gamma_{\rm out}\left(1\pm \hat{v}_{s}v_{\rm out}\cos\xi\right)\right]^{-1},
\end{aligned}
\label{epsiloninplunging}
\end{equation}
where $v_{\rm out}$ represents the outflow velocity and $\gamma_{\rm out}$ denotes its corresponding Lorentz factor. These quantities are defined by the magnetization parameter $\sigma$ as
\begin{equation}
v_{\rm out}=\sqrt{\frac{\sigma}{1+\sigma}},\qquad\gamma_{\rm out}=\sqrt{1+\sigma}.
\label{lorentzplunging}
\end{equation}
Similar to the circular orbit case, we now investigate the allowed energy extraction region in $r-a$ plane for the plunging region. The corresponding results are presented in Figs. {\bf\ref{fig12}-\ref{fig15}}, The red, blue, purple and black lines represents the event horizon, ergosphere, photon sphere radius and ISCO, respectively. From left to right $\sigma=100,~30,~10,~3$, respectively.
From Fig. {\bf\ref{fig12} (a)}, it can be observed that the allowed energy extraction region expands with increasing values of $\sigma$. Comparing this result with Fig. {\bf\ref{fig4} (a)}, corresponding to the circular orbit case $r>r_I$, we find that the allowed regions are identical outside the ISCO. However, for $r<r_I$, the allowed energy extraction region in the plunging region is significantly larger than that in the circular orbit case. In particular, the minimum black hole spin required for energy extraction is substantially lower in the plunging region. Furthermore, the position of the reconnection layer extends farther inward in the plunging region, in agreement with Ref. \cite{shen2024energy}. For example, when $\sigma=100$, the minimum spin required for energy extraction is $0.86$ in the circular orbit case, as shown in Fig. {\bf\ref{fig4} (a)}, whereas it decreases to approximately $0.25$ in the plunging region. A comparison of the three panels of Fig. {\bf\ref{fig12}} further shows that, as the value of $\rho_c$ increases, the minimum spin required for energy extraction increases. Specifically, the minimum allowed spin increases from $0.25$ in Fig. {\bf\ref{fig12} (a)} to $0.30$ in Fig. {\bf\ref{fig12} (c)}, making energy extraction possible even for more slowly rotating black holes.
From Fig. {\bf\ref{fig13}}, it can be observed that the minimum black hole spin required for energy extraction increases with increasing values of  $R_s$, which is consistent with the behaviour observed in the circular orbit case. Therefore, we conclude that both  parameters, $\rho_c $ and $R_s$, influence the minimum spin required for energy extraction. Nevertheless, the minimum allowed spin in the plunging region remains significantly lower than that in the circular orbit case. For example, at $\sigma=100$, the minimum spin required for energy extraction in the plunging region is approximately $0.25$, whereas the corresponding value in the circular orbit case Fig. {\bf\ref{fig5} (c)} is $0.90$. This clearly highlights the enhanced capability of the plunging region to support energy extraction from slowly rotating black holes.
From Fig. {\bf\ref{fig14}}, it can be observed that the shape of the allowed energy extraction region changes with the azimuthal angle $\xi$. This behavior is a characteristic feature of the plunging region, where the allowed region depends on the azimuthal angle. Our results is consistent with those reported in Ref. \cite{shen2024energy}.

Following Fig. {\bf\ref{fig7}}, now we plot the allowed region for fixed values of parameters $\rho_c$ and $R_s$ in Fig. {\bf\ref{fig15}}. Here we can see that the minimum allowed spin for energy extraction is $0.25$, which is lower than the circular orbit case.

\subsection{Power of Energy Extraction in the Plunging Region}

Similar to the circular orbit case, we now investigate the energy extraction power in the plunging region, still using Eq. (\ref{powerext}). We compare the extracted power for both the plunging region and the circular orbit case in the region $r<r_{I}$. For this analysis, we choose the parameter values $\rho_c=0.25,~R_s=0.1,~a=0.9,~\xi=\pi/12$, and $\sigma=100$, here all values of $r$ are less than ISCO.
\begin{figure}[H]
\centering 
\includegraphics[width=0.5\linewidth]{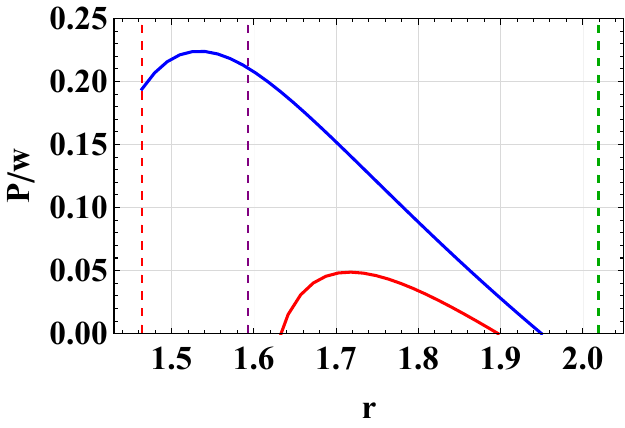}
\caption{Red dashed, purple dashed, green dashed lines represent the event horizon, photon sphere
radius, ergosphere, respectively. Red solid line represents the circular orbit power and blue solid
line represents the plunging region power.}
\label{fig16}
\end{figure}
From Fig. {\bf\ref{fig16}}, it can be observed that the magnetic reconnection power in the plunging region is greater than that in the circular orbit case. Finally, we compare the magnetic reconnection power in the plunging region with that of the Blandford–Znajek mechanism using Eq. (\ref{powerratio}). For this comparison, we use the same parameter values as those adopted in the circular orbit case, namely $a=0.98,~r=1.4,~\sigma=7,~\rho_c=0.05$, and $R_s=0.2$. Substituting these parameter values into the relevant expressions, we obtain $r_I=1.74442$, which satisfies the plunging region condition $r<r_I$. The resulting power ratio is found to be $27.5438$, which is greater than the value $4.08458$ obtained for the circular orbit case. These results demonstrate that the magnetic reconnection power in the plunging region exceeds that of both the circular orbit case and the Blandford–Znajek mechanism. We now consider the low spin case based on Fig. {\bf\ref{fig12} (c)}, for which the parameter values are $\rho_c=0.05,~R_s=0.1,~\xi=\pi/12$, and the black hole spin is $a=0.25$. In this case, the photon sphere radius lies above the ergosphere; therefore, Eq. (\ref{crossarea}) is no longer applicable. Instead, we approximate the inflowing plasma cross-sectional area as
\begin{equation}
 A_{\rm in}\sim \left(r_{\rm E}^{2}-r_{+}^{2}\right),   
\end{equation}
 Using the parameter values $\sigma=100,~a=0.25,~r=0.5$, and $\xi=\pi/12$, we obtain a power ratio of approximately $0.314602$, which is relatively small. This result is physically reasonable because achieving energy extraction from such a low spin black hole requires a large value of $\sigma$. According to Eq. (\ref{powerratio}), $\sigma$ appears in the denominator of the power ratio, thereby reducing its magnitude. Nevertheless, the ability to extract energy through the magnetic reconnection mechanism from a black hole with a spin as low as $0.25$ is a significant result.
 
\section{CONCLUSION}

In this work, we have investigated the extraction of rotational energy through the Comisso–Asenjo magnetic reconnection mechanism in the spacetime of a rotating black hole surrounded by CDM halo. Using the Kerr-like black hole solution immersed in a CDM halo, we analyzed the influence of the DM parameters on the spacetime geometry and on the efficiency of magnetic reconnection in both the circular orbit and plunging plasma regions. We first examined the geometrical properties of the spacetime and showed that the CDM halo modifies the event horizon, photon sphere, and ergosphere, thereby changing the region in which magnetic reconnection can operate. The conditions required for energy extraction were then analyzed in terms of the energy-at-infinity of the accelerated and decelerated plasma components. Our results demonstrate that the magnetic reconnection mechanism remains efficient in the presence of a CDM halo and that the allowed parameter space depends sensitively on the characteristic density parameter $\rho_c$, the halo scale radius $R_s$, the plasma magnetization $\sigma$, and the magnetic field orientation angle.

For circular plasma motion, we found that energy extraction is possible for black holes with spin as low as $a\simeq 0.86$. Although increasing $\rho_c$ and $R_s$ reduces the extracted power and efficiency for a fixed spin, these parameters simultaneously lower the minimum spin required for magnetic reconnection to occur, allowing the mechanism to operate around more moderately rotating black holes. Furthermore, the extracted power increases with increasing plasma magnetization and exceeds the corresponding Blandford–Znajek power over the considered parameter space, confirming the high efficiency of the Comisso–Asenjo mechanism. We further extended the analysis to the plunging plasma region inside the ISCO, where the radial motion of the plasma becomes significant. In this region, the allowed parameter space for energy extraction becomes considerably larger than in the circular orbit case. In particular, the minimum spin required for successful energy extraction decreases to approximately $a \simeq 0.25$, demonstrating that magnetic reconnection can efficiently extract rotational energy even from slowly rotating black holes. We also showed that the magnetic reconnection power in the plunging region is larger than that obtained for circular orbits and, for suitable parameter choices, substantially exceeds the Blandford–Znajek prediction.

Overall, our results indicate that the presence of a CDM halos has a significant impact on magnetic reconnection around rotating black holes by extending the range of black hole spins capable of powering this process. These findings suggest that magnetic reconnection remains a robust and highly efficient mechanism for extracting rotational energy in realistic astrophysical environments where supermassive black holes are expected to reside within DM halos. The present analysis may provide further insight into the origin of high energy astrophysical phenomena associated with accreting black holes and offers a useful framework for future investigations of magnetic reconnection in more realistic black hole environments, including different DM distributions, magnetized accretion flows, and observational constraints from gravitational-wave and EHT measurements.

\end{document}